\documentclass[11pt,a4paper]{article}

\usepackage[utf8]{inputenc}
\usepackage[english]{babel}
\usepackage[a4paper, total={6in, 9in}]{geometry}
\usepackage{amsfonts}
\usepackage{amsmath}
\usepackage{amssymb}
\usepackage{import}
\usepackage{cancel}
\usepackage{feynmp-auto}
\usepackage{graphicx}
\usepackage{slashed}
\usepackage{array}
\usepackage{mathrsfs}
\usepackage{lineno}
\usepackage{float}
\restylefloat{table}
\usepackage{verbatim}
\usepackage{xcolor}
\usepackage{color, colortbl}
\usepackage{upgreek}
\usepackage{comment}
\usepackage{multirow}
\usepackage{enumitem}
\usepackage{mathtools}
\usepackage[vcentermath]{youngtab}
\usepackage{jheppub}
\usepackage[T1]{fontenc} 
\usepackage{subcaption}
\usepackage{longtable} 
\usepackage[compat=1.1.0]{tikz-feynhand}	
\usepackage{tikz-feynman}
\tikzfeynmanset{compat=1.1.0}
\usepackage{feynmp}
\usepackage{tikzsymbols}
\usepackage{array}
\usepackage{pifont} 
\usepackage{soul} 
\usepackage{ytableau} 

\newcommand{\T}{\tilde}

\newcommand{\luv}{\Lambda}
\newcommand{\mh}{M_{\text{H}}}

\newcommand{\+}{\oplus}

\newcommand{\q}{\mathcal{Q}}
\definecolor{Gray}{gray}{0.9}
\definecolor{LightCyan}{rgb}{0.88,1,1}
\definecolor{maroon}{cmyk}{0,0.87,0.68,0.32}
\newcolumntype{P}[1]{>{\centering\arraybackslash}p{#1}} 

\newcommand{\ov}{\overline}

\newcommand{\diag}{\text{diag}}

\newcommand{\sun}{SU(N)_{\mathrm{DC}}}

\newcommand{\ldc}{\Lambda_{\mathrm{DC}}}

\newcommand{\sundc}{SU(N)_{\mathrm{DC}}}
\newcommand{\sondc}{SO(N)_{\mathrm{DC}}}

\newcommand{\ndc}{N_{\mathrm{DC}}}
\newcommand{\ndf}{N_{\mathrm{DF}}}

\newcommand{\mL}{m}
\newcommand{\mH}{M}
\newcommand{\mgut}{M_{\text{GUT}}}
\newcommand{\vgut}{v_\text{GUT}}
\newcommand{\sgut}{s_\text{GUT}}
\newcommand{\Sgut}{\sigma_\text{GUT}}
\newcommand{\mDM}{m_{DM}}
\newcommand{\GDC}{{\cal G}_{\text{DC}}}
\newcommand{\GU}{{\cal G}_U}
\newcommand{\GSM}{{\cal G}_{\text{SM}}}
\newcommand{\mpiD}{m_{\pi_D}}

\title{A GUT Framework for \\ Accidental Composite Dark Matter}
\author[a]{Salvatore Bottaro,}
\author[b, \, c]{Roberto Contino,}
\author[d]{Sonali Verma}
\affiliation[a]{Raymond and Beverly Sackler School of Physics and Astronomy,  Tel-Aviv, Israel}
\affiliation[b]{Dipartimento di Fisica, Sapienza Universit\`a di Roma,  Italy}
\affiliation[c]{Istituto Nazionale di Fisica Nucleare (INFN), Sezione di Roma, Italy}
\affiliation[d]{Service de Physique Th\'eorique, Universit\'e Libre de Bruxelles, Belgium}

\emailAdd{salvatoreb@tauex.tau.ac.il, roberto.contino@uniroma1.it, sonali.verma@ulb.be}
\date{February 2021}
\abstract{We study and classify $SU(5)$-GUT completions of accidental composite dark matter models. These theories postulate new vectorlike confining dark color dynamics and give an accidentally stable baryonic dark matter candidate. In realistic theories, dark fermion $SU(5)$ irreps split into light dark quarks, whose bound states include the dark matter, and their much heavier GUT partners. 
A simple analysis shows that such a mass hierarchy requires a fine tuning of parameters and thus implies a naturalness problem.
We select theories requiring that all dangerous metastable states decay before the onset of nucleosynthesis through higher-dimensional operators generated at the GUT scale or at the mass scale of dark quark GUT partners.
Demanding Standard Model gauge coupling unification puts severe constraints on the landscape of viable theories.
Under the assumption of an approximately degenerate spectrum of dark quark GUT partners, we find that only one model gives precision unification.}

\begin{document}
\maketitle

\section{Introduction}
The trajectories of the Standard Model (SM) gauge couplings, when extrapolated using Renormalization Group Equations (RGE), come close to each other in value at an energy of order $\sim 10^{14} \,\mathrm{GeV}$.
This observation can hint at the intriguing possibility that the SM gauge group unifies under a Grand Unified Theory (GUT). The simplest scenario of non-supersymmetric minimal $SU(5)$ GUT, where the SM matter is unified in the $\bar{5}$ and $10$ representations of $SU(5)$, has been already excluded~\cite{Super-Kamiokande:2016exg}. The GUT gauge bosons mediate proton decay, and the proton lifetime bounds set a lower constraint on the unification scale $\mgut \gtrsim 10^{15} \,\mathrm{GeV}$. 
The evolution of the SM gauge couplings is sensitive to any new fields with quantum numbers under the SM gauge group. Unification can be thus used as a guiding principle to parametrize any new physics (NP) content (see \cite{Giudice:2004tc} for example).

The SM can be further considered paradigmatic for its other remarkable properties. In particular, global symmetries in the SM arise accidentally in the infrared and explain conservation laws violated only by UV-suppressed higher-dimensional operators. A minimal solution to the cosmological stability of dark matter (DM) can be found using accidental symmetries, in analogy with the proton stability arising from baryon number conservation in the SM. Theories of this kind are free from ad-hoc symmetries (like R-parity or $Z_2$), which are imposed by hand in most DM models.

A simple way to get an accidentally long-lived DM state is to extend the SM gauge group with a new confining gauge force.~\footnote{For theories of accidental dark matter without new gauge groups see Ref.~\cite{Cirelli:2005uq} and Ref.~\cite{Cheung:2015mea}.}
In this paper, we investigate the possibility that the dark sector content comprises new dark quarks transforming as real or vectorlike representations under the SM gauge group and the new dark gauge group. This framework, first introduced in \cite{Kilic:2009mi} and dubbed \textit{vectorlike confinement}, allows mass terms for the dark fermions and minimizes the impact on electroweak precision observables.
Accidental DM theories of this kind were previously considered in~\cite{Antipin:2015xia, Mitridate:2017oky, Contino:2018crt}.
Adding chiral dark fermions, in contrast, is much trickier due to non-trivial anomaly cancellation (See Ref.~\cite{Contino:2020god} for example). In this work, we concentrate on baryonic DM candidates, which arise naturally from the strong dynamics and are generically more robust than mesonic DM candidates.

There is a large number of theories and models that can give rise to accidental composite DM states. However, we find that \textit{demanding that any new particles mitigate (and not worsen) gauge coupling unification in the SM puts severe restrictions on these constructions}. As a matter of fact, very few models pass our criteria for gauge coupling unification (see Sec.~\ref{sec:unification}).

The main motivation of this work is to systematically describe the $SU(5)$-GUT completion of accidental composite dark matter models.
The DM sector comprises (light) dark fermions which transform in the fundamental representation of the dark color gauge group $\mathcal{G}_{DC}$, identified with $\sundc$ or $\sondc$, and have masses below the dark confinement scale, $\ldc$.
These dark fermions come as $SU(5)$ fragments and in a consistent $SU(5)$-GUT theory must have their own (heavy) GUT partners. We are thus led to ask the following question:
\textit{how relevant are the dark fermion GUT partners for the dark matter model building, and how do they impact cosmology?}
We systematically address this question in this work and in a forthcoming paper~\cite{BCV}.

The dark sector gauge symmetry in the UV is $\mathcal{G}_U \times \mathcal{G}_{\text{DC}}$, with $\mathcal{G}_U$ being the unified group $SU(5)$. At the scale $\mgut$, the unified group breaks spontaneously to the SM group $\mathcal{G}_{\text{SM}} = SU(3)\times SU(2)\times U(1)$. From $\mgut$ down to the dark confinement scale $\Lambda_{\text{DC}}$, the gauge symmetry of the theory is thus $\mathcal{G}_{\text{SM}} \times \mathcal{G}_{\text{DC}}$.  The requirement of SM gauge coupling unification sets the mass scale of the GUT partners and implies the following hierarchy of mass scales: $$\mL \lesssim \ldc < \mH < \mgut.$$ Here $\mL$ is the mass of the light dark fermions and $\mH$ is the mass of their GUT partners.

Before describing our dark sector, we would like to discuss previous works that have performed studies similar to ours. Apart from Ref.~\cite{Antipin:2015xia}, we are not aware of other works in the literature that consider accidentally stable dark matter models in the context of $SU(5)$ unification.~\footnote{Axion models have been constructed in the context of $SU(5)$ unification where the Peccei-Quinn symmetry is accidental, see for example~\cite{Vecchi:2021shj,Contino:2021ayn}.}
In \cite{Antipin:2015xia}, the authors perform a systematic classification of strongly-coupled vectorlike theories that give an accidentally stable DM candidate. In this paper, we extend the work of Ref.~\cite{Antipin:2015xia} as follows:
\begin{itemize}
\item While in Ref.~\cite{Antipin:2015xia} dark quarks are assumed to come in $SU(5)$ fragments, the classification is done by focusing on the low-energy DM sector content. Models are characterized as either \textit{golden class} (models where all dark species symmetries are broken at the renormalizable level), or \textit{silver class}  (models which can be made viable either by requiring higher-dimensional operators to break dark species symmetries or via additional model building).
  Our classification, instead, is done from an $SU(5)$-GUT perspective. We require that unwanted species number symmetries of the dark matter sector are broken by higher-dimensional operators generated by the GUT dynamics or at the mass scale of dark quark GUT partners.
We reproduce all the golden class models of Ref.~\cite{Antipin:2015xia} but obtain a different selection of silver class models.
We describe this analysis in Sec.~\ref{sec:classification}.
\item In Ref.~\cite{Antipin:2015xia} the question of gauge coupling unification relies on perfect unification (where the three couplings meet at a point in the $(\log \mu, \alpha_i^{-1})$ plane), and the unification analysis is restricted to only three viable models. Our unification criteria discussed in Sec.~\ref{sec:unification} are more relaxed, and we also consider SM extensions that simply improve upon the unification found with the minimal SM content.
\item  When discussing unification, Ref.~\cite{Antipin:2015xia} assumes a common mass scale for the GUT partners. We also consider some scenarios with a mass splitting between the GUT partners, details of this analysis can be found in Sec.~\ref{section:nondegmass}.
\item We analyze the issue of producing a natural hierarchy between the mass of the light dark fermions and that of their GUT partners, $\mL \lesssim \ldc \ll \mH$. We find that such hierarchy requires fine tuning, and thus implies a problem of naturalness in these theories.
Details can be found in Sec.~\ref{sec:mass_split}.
\end{itemize}

There have been previous works in the literature that aim at solving the DM problem along with gauge coupling unification, see for example~\cite{Ibe:2009gt, Aizawa:2014iea, Mahbubani:2005pt, Harigaya:2016vda}. None of these models, however, utilize accidental symmetries to stabilize the DM candidate.
The model studied in~\cite{Harigaya:2016vda} can, in fact, be considered a particular realization, with an $SU(3)$ dark color group, of the class of theories studied in this work. The mass of light dark quarks, hence of the lightest dark baryon, is assumed to be much larger than the lower bound set by direct detection experiments on DM candidates with non-zero hypercharge (the bound from LUX used in~\cite{Harigaya:2016vda} was $\sim 10^7\,$GeV, while the current bound from LZ is  $\sim 10^9\,$GeV). This allowed the authors to consider a DM sector where the only light dark quark, $Q$, transforms as a $(3,2)_{1/6}$ under the SM group, and to identify the DM candidate with the EM-neutral component of the $QQQ$ baryon multiplet. Due to its extremely high mass, the DM is not accidentally stable (as $D=6$ dark baryon number violating operators would mediate a too fast decay), and its abundance was assumed to be produced before the end of reheating.
Therefore, although the construction of Ref.~\cite{Harigaya:2016vda}  and the models considered in this paper belong to the same theoretical framework, they lead to rather different phenomenologies and cosmologies.

Our paper is organized as follows. Section~\ref{sec:symmetries} contains an analysis of the accidental symmetries in our theories and of the higher-dimensional operators that break them. We outline our classification of candidate models in Sec.~\ref{sec:classification}, and report its results in Appendix~\ref{appendix:models_list}.
Our analysis of SM gauge coupling unification is illustrated in Sec.~\ref{sec:unification} for each of these models.
We address the issue of mass splitting between dark fermions and their GUT partners in Sec.~\ref{sec:mass_split}, and present a detailed derivation of the tree-level mass spectrum of the $Q\oplus \tilde D$ model in Appendix~\ref{appendix:splitting}.
We summarize and conclude in Sec.~\ref{sec:summary_cosmo}.

\section{Accidental Symmetries}
\label{sec:symmetries}

Consider a minimal (non-supersymmetric) $SU(5)$ Grand Unification scenario where SM fermions are embedded into chiral $\bar 5$ and $10$ representations and GUT scalars transform as $5$ and $24$. We extend such theory by postulating a new confining gauge group $\mathcal{G}_{\text{DC}} = \sundc$ or $\sondc$  (\textit{dark color}),~\footnote{We do not consider $Sp(N_{DC})$ gauge theories since they do not have stable dark baryons~\cite{Witten:1983tx}.} and introducing new fermions (\textit{dark fermions}) which transform as fundamentals under $\mathcal{G}_{\text{DC}}$ and as irreps $R$ of ${\cal G}_U = SU(5)$. For simplicity, we will restrict our analysis to irreps $R$ with rank two or lower: $R = 1, 5, 10, 15, 24$. In general, $(\square, R)$ is a complex representation of $\GDC\times\GU$ and the corresponding dark fermion field is defined to be a Dirac spinor; in that case, dark fermions are vectorlike irreps of $\GDC\times\GU$. Whenever $(\square, R)$ is real (this happens if $\GDC = \sondc$ and $R = 1$ or 24), the dark fermion field is defined to be a Majorana spinor; in that case, dark fermions are real irreps of $\GDC\times\GU$. No scalars colored under $\GDC$ (i.e. dark scalars) are introduced for simplicity (see Refs.~\cite{Buttazzo:2019iwr,Palmisano:2024mxj} for examples of accidental DM theories with dark scalars).

At the scale, $\mgut$, the unified group ${\cal G}_U$ is spontaneously broken to $\mathcal{G}_{SM}$ by the vev of the scalar 24, and dark fermions break up into the $SU(5)$ fragments listed in Tab.~\ref{tab:notation}.
We use the symbol $\q_r$ to indicate a generic fragment, where the index $r$ labels different \textit{dark species}, i.e. SM irreps.
Individual SM irreps are instead denoted by following the notation of Ref.~\cite{Antipin:2015xia}, see Tab.~\ref{tab:notation}.
\begin{table}
\begin{center}
\begin{tabular}{ |c|c|c| } 
\hline
\rowcolor[gray]{.9}
$\q_r$ & $SU(5)$ & SM Quantum Numbers\\[0.1cm]
\hline
\hline
 $N$ & 1 & $(1,1,0)$  \\[0.1cm]
 $D \oplus L$ & $\bar{5}$ & $(\bar{3},1,1/3) \oplus (1,2,-1/2)$  \\[0.1cm]
 $U \oplus E \oplus Q $ & 10  & $(\bar{3},1,-2/3) \oplus (1,1,1) \oplus (3,2,1/6)$  \\[0.1cm]
 $Q \oplus T \oplus S$ & 15 & $(3,2,1/6) \oplus (1,3,1) \oplus (6,1,-2/3)$  \\[0.1cm]
 $V \oplus G \oplus X \oplus N$ & 24 & $(1,3,0) \oplus (8,1,0) \oplus [ (\bar{3},2,5/6) \oplus ({3},2, -5/6)] \oplus (1,1,0)$\\[0.1cm]
\hline
\end{tabular}
\end{center}
\caption{Dark fermions $\q_r$ and their corresponding quantum numbers under the SM gauge group along with the $SU(5)$ representations embedding them. The field $X$ is used to denote the reducible representation $(\bar{3},2,5/6) \oplus ({3},2, -5/6)$.}
\label{tab:notation}
\end{table}
At energies much below the scale $\mgut$, the dark sector Lagrangian has the following form at the renormalizable level:~\footnote{In this paper we use a 4-component notation for spinor fields.}
\begin{equation}
  \label{eq:dark_lagrangian}
\begin{split}
  \mathcal{L}_{DS}^{(4)} = 
  & - \frac{1}{4 g_{DC}^2} (\mathcal{G}^A_{\mu \nu})^2 + \frac{\theta_{DC}}{32\pi^2} \mathcal{G}^A_{\mu \nu} \mathcal{\tilde G}^{A,\mu \nu}+ \bar{\q}_r (i \slashed{D} - m_{\q_r}) \q_r \\[0.2cm]
  & -  \bar{\q}_r H \left[ (y_L)_{rs} P_L + (y_R)_{rs} P_R \right] \!\q_s + \text{h.c.}
\end{split}
\end{equation}
Here $\mathcal{G}^A_{\mu \nu}$ is the dark gluon field strength, $g_{\text{DC}}$ is the dark gauge coupling, $P_{L,R}$ are projectors over left- and right-handed chiralities, and a sum over the indices $A$, $r,s$ has been left understood.
Depending on the SM quantum numbers of the dark fermions $\q_r$, Yukawa couplings with the Higgs boson can be written, as shown in Eq.~(\ref{eq:dark_lagrangian}).

The dark color group ${\cal G}_{\text{DC}}$ is assumed to confine at a scale $\ldc \ll \mgut$, thus forming a spectrum of dark baryons and mesons. For a detailed characterization of these hadrons and their quantum numbers, we refer the reader to Ref.~\cite{Antipin:2015xia}. In the following, we will focus on the accidental symmetries that emerge at low energy. They are different for $\sundc$ and $\sondc$ theories, so we will analyze the two cases separately.

In theories with $\GDC = \sundc$, where dark fermion fields are Dirac spinors, the Lagrangian (\ref{eq:dark_lagrangian}) is invariant under a common phase transformation of all the $\q_r$. This is the dark baryon number symmetry,  $U(1)_{\text{DB}}$, which therefore emerges as an accidental symmetry of the dark sector at low energy and can make the lightest baryon cosmologically stable. 
Besides dark baryon number, $\mathcal{L}_{DS}^{(4)}$ can have additional accidental symmetries, such as species number symmetries and dark G-parity,  which can lead to the stability of other bound states. Species number symmetries correspond to the relative phase rotations of dark species, while dark G-parity was defined  in~\cite{Bai:2010qg} as a modified charge conjugation acting (only) on dark fields. Gauge interactions of the dark quarks are invariant under dark G-parity if the $\q_r$ transform as real or pseudoreal representations of the SM gauge group. This restricts the possible dark species to $V$, $G$ and $N$ (see Tab.~\ref{tab:notation}), although in practice $\sundc$ theories with $G$ dark fermions violate the perturbativity requirement formulated in Sec.~\ref{sec:classification} (they have $SU(3)$ Landau poles below $\mgut$) and as such will be discarded in our analysis. It is easy to see that dark G-parity is an accidental invariance of the GUT theory as well if dark fermions transform in real irreps of $\GU$, that is if $R = 1, 24$.

If $\GDC = \sondc$, dark baryon number is anomalous and explicitly broken by Majorana mass terms, hence there is no distinction between baryons and antibaryons. On the other hand, Eq.~(\ref{eq:dark_lagrangian}) turns out to be invariant under $O(N)_{\text{DC}}$ global transformations, i.e. the symmetry $\mathbb{Z}_2 = O(N)_{\text{DC}}/\sondc$ is accidental~\cite{Antipin:2015xia}. Since dark baryons and mesons are respectively odd and even under such parity, the lightest dark baryon is accidentally stable.
Dark species number symmetries can still be defined for dark fermions in complex SM irreps, whose fields are Dirac spinors. For Majorana fermions, one can define species parities under which a given species is odd, while the others are even.~\footnote{The $\mathbb{Z}_2$ transformation under which all dark quark fields change sign, $\q_r \to - \q_r$, is not a new symmetry: for $N_{\text{DC}}$ even, it is an element of the gauge group $\sondc$, while for $N_{\text{DC}}$ odd it coincides with $O(N)_{\text{DC}}/\sondc$. It is the residual anomaly-free discrete symmetry, contained in the classical $U(1)$ global symmetry, that is preserved by the dark color condensate and by the fermionic mass terms.}
%
Discrete and continuous species symmetries leave the Lagrangian  (\ref{eq:dark_lagrangian}) invariant in the absence of Yukawa interactions. Finally, dark G-parity is not useful for $\sondc$ theories, because fields transforming in real or pseudoreal irreps of $\GSM$ are Majorana spinors, and as such they are invariant under (dark) charge conjugation. This implies that all dark mesons made of such irreps are even under dark G-parity.

Accidental symmetries can be used to obtain a cosmologically stable DM candidate. At the same time, their presence can be a potential problem if it leads to metastable states with dangerous SM quantum numbers. Selecting viable models thus requires a careful analysis of accidental symmetries.
In doing so, as explained in the following, one finds that the spectrum of dark fermions must be split: some dark species must be light (with mass $\mL \lesssim \ldc$) to form the DM candidate bound state, while others must have a large mass~$\mH$, with $\ldc \ll \mH \lesssim\mgut$, in order to decay before BBN.  We will refer to these sets of light and heavy dark fermions as respectively the \textit{DM sector} and the sector of dark fermion \textit{GUT partners}.

\subsection{Viable Accidental DM Candidates}
\label{sec:high_dim_Os}

There are two kinds of accidental DM candidates in our class of theories: dark baryons and dark pions, respectively stabilized by dark baryon number, and by species numbers or G-parity. As already noticed previously~\cite{Bai:2010qg,Antipin:2015xia} and argued below, dark baryons are more robust candidates compared to dark pions and will account for the whole DM abundance in the models studied in this work.

When dealing with accidental symmetries, it is crucial to determine at which level they are violated, i.e. to identify the
symmetry-breaking operators with the smallest dimension. 
Let ${\cal O}$ be one such operator and let us denote its dimension with $D$. If generated at the scale $\Lambda$ with a coefficient $k$,
\begin{equation}
\label{eq:high_dimO} 
\Delta {\cal L} =  \frac{k}{\luv^{D-4}} \mathcal{O} \, ,
\end{equation}
such an operator induces the decay of the lightest particle charged under the accidental symmetry with a rate of order~\footnote{The factor $(\ndc/16\pi^2)$ comes from the matrix element of ${\cal O}$ between the decaying particle and the vacuum, and it follows from a large-$\ndc$ counting. Our estimate assumes a two-body final state. An additional suppression factor should be included for decays that involve more than two final states or that arise at a higher order in the loop expansion. An extra suppression compared to the estimate of Eq.~(\ref{eq:gamma}) also arises if the decay involves the virtual propagation of SM fields with masses much higher than $m_{BS}$.}
\begin{equation}
\label{eq:gamma}
\Gamma \sim \frac{k^2}{8\pi}\frac{\ndc}{16\pi^2}m_{BS} \left(\frac{m_{BS}}{\luv}\right)^{2D-8}\, ,
\end{equation}
where $m_{BS}$ is the mass of the particle.

The lightest state charged under the accidental symmetry can be an acceptable DM candidate if its lifetime satisfies the bounds set by indirect detection DM searches~\cite{Cohen:2016uyg,Ando:2015qda,Cirelli:2012ut,Facchinetti:2023slb,Wadekar:2021qae, Arguelles:2019boy},
\begin{equation}\label{eq:DMtau}
\tau_{\text{DM}} \gtrsim \left( 10^{25} - 10^{28} \right) \mathrm{s}\, ,
\end{equation}
in the range of DM masses $\mDM = (10^{-3} - 10^5) \,$GeV and for a variety of decay channels.
By using the estimate of Eq.~(\ref{eq:gamma}), these constraints translate into an upper bound on the DM mass $\mDM$ as a function of $D$ and $\luv$. In particular, identifying $\Lambda$ with the reduced Planck scale $M_P= 2.4 \times 10^{18}\,$GeV and setting $k^2 \ndc =1$ leads to the following bounds from dimension-5 and dimension-6 operators:
\begin{align}
    \label{eq:DM_D5}
  \text{For $D = 5$:} \quad \quad \mDM &\lesssim 1.5\,\mathrm{MeV}\, , \\[0.2cm]
    \label{eq:DM_D6}
  \text{For $D = 6$:} \quad \quad \mDM &\lesssim 100\,  \mathrm{TeV} \, .
\end{align}
The bound of Eq.~(\ref{eq:DM_D5}) is set by data on the radiative cooling rate of gas in dwarf galaxies~\cite{Wadekar:2021qae} and applies to DM decays into $e^+e^-$ mediated by an EM dipole operator $\bar\q \sigma^{\mu\nu}\q F_{\mu\nu}$. The bound of Eq.~(\ref{eq:DM_D6}) instead comes from gamma-ray data from Fermi (see for example~\cite{Cohen:2016uyg}) and the neutrino flux from IceCube~\cite{Arguelles:2019boy}.

Since in our theories, dark baryon number is broken by operators with dimension 6 or higher, from Eq.~(\ref{eq:DM_D6}) it follows that dark baryons much heavier than the electroweak scale can be good accidental DM candidates. Species numbers and G-parity, on the other hand, are generally broken at the $D=5$ level in vectorlike theories~\cite{Bai:2010qg,Antipin:2015xia}~\footnote{The stability of dark mesons can be enhanced in theories with a chiral gauge group, see~\cite{Contino:2020god}.}; Eq.~(\ref{eq:DM_D5}) thus suggests that constructing a viable theory where the DM abundance comes even partly from accidentally stable dark pions is not straightforward. Indeed, the mass square of a pseudo Nambu-Goldstone boson (NGB) can be estimated by
\begin{equation}
\label{eq:mpiD}
\mpiD^2 = c_1 \frac{g_{SM}^2}{16\pi^2} \ldc^2+ c_2 m \ldc,
\end{equation}
where $g_{SM}$ is an SM coupling and $c_1$, $c_2$ are $\mathcal{O}(1)$ coefficients.
The first term arises from SM loops and implies that  $\mpiD$ can be taken much smaller than the electroweak scale only at the price of assuming correspondingly low values of the dark scale $\ldc$. These are excluded by collider searches as long as the dark sector comprises SM-charged constituents.
One is thus left to consider SM singlet dark pions, for which $c_1=0$. Such particles can be much lighter than the electroweak scale for $m \ll \ldc$, and could in principle play the role of accidental DM candidates.
We leave the study of theories of this kind to a future work and focus instead on dark baryon candidates in this paper.

To further substantiate our analysis of dark pions as accidental DM candidates, we now show that their stabilizing symmetry is always broken at the $D=5$ level in our theories.
First, notice that dark matter candidates with mass smaller than $\sim 9\, y^2\times 10^{9}\,$GeV are excluded by direct detection bounds~\cite{LZ:2022lsv} if they have non-vanishing hypercharge $y$ and thus couple to the $Z$ boson. We must therefore restrict our analysis to DM candidates with zero hypercharge; these have integer weak isospin since they have zero electromagnetic (EM) charge.
We also assume that the DM is a singlet under SM color to avoid the strong experimental bounds on that scenario (see~\cite{DeLuca:2018mzn} for an example of theories with colored DM).
If one excludes SM singlets, the only dark pions that
fulfill these requirements and have a stabilizing accidental symmetry are~\footnote{Here and in the following, we characterize dark hadrons by means of the local operators that excite them from the vacuum.}
\begin{equation}
   \label{eq:dm_mesons}
  \begin{split}
    \bar{T}E, \quad \bar{V}N, \quad \bar{V}V \quad \text{and} \ \, \bar{G}G \quad & \text{in $\sundc$ theories,} \\[0.1cm]
    \bar{T}E \quad \text{and} \ \, \bar VN \quad & \text{in $\sondc$ theories.}
    \end{split}
\end{equation}
While $\bar{T}E$, $\bar{V}N$ are stabilized by species symmetries, $\bar{V}V$ and $\bar{G}G$ can have odd G-parity.  As anticipated, $\sundc$ theories with $G$ dark quarks do not satisfy the perturbativity requirement of Sec.~\ref{sec:classification} and are thus discarded in our analysis. This eliminates $\bar GG$ states as potential DM candidates. The remaining states in Eq.~(\ref{eq:dm_mesons}) all transform as electroweak triplets with zero hypercharge, and their electromagnetically neutral component could act as DM. Their stabilizing symmetries can however all be broken by $D=5$ operators such as
\begin{equation}
  \label{eq:D5op}
\begin{split}
& \epsilon^{ijk} \bar V^j V^k (H^\dagger \sigma^i H) , \quad \bar V \sigma^{\mu\nu} V B_{\mu\nu} , \\[0.1cm]
& \bar V^i N (H^\dagger \sigma^i H) , \quad \bar V^i \sigma^{\mu\nu} \! N\,  W^i_{\mu\nu}, \\[0.1cm]
& \bar T^i E (H^\dagger \sigma^i H) , \quad \bar T^ i \sigma^{\mu\nu} \! E \, W^i_{\mu\nu} .
\end{split}
\end{equation}
This implies that none of the dark pions of Eq.~(\ref{eq:dm_mesons}) can play the role of accidental DM. Dark pions that are SM singlets can also be accidentally stable due to a species symmetry (this requires to have several flavors of dark quark singlets), though this symmetry will also be broken at the $D=5$ level. As said earlier, we do not consider such a possibility here.  Therefore, in our theories, the DM candidate will be identified with the lightest accidentally stable dark baryon.

Besides their accidental stability, another remarkable feature of dark baryons is that they can reproduce the observed DM abundance through thermal freeze-out. Their dominant annihilation channel is into dark pions, with a thermally averaged cross-section of order $\langle\sigma v\rangle \sim \pi/\ldc^2$. This implies a correct DM density for $\ldc \sim 50\,$TeV, which corresponds to $\mDM \sim \ndc\, 50\,$TeV and is thus compatible with the bound of Eq.~(\ref{eq:DM_D6}) for not too large values of $\ndc$.
Thermal dark baryons are therefore good accidental DM candidates. It is also interesting to notice that future indirect detection experiments will have the sensitivity to observe DM decays if these are mediated by $D=6$ dark baryon-violating operators (this is because DM masses of order $100\,$TeV saturate the bound of Eq.~(\ref{eq:DM_D6})).

\subsection{Other long-lived bound states}
\label{subsec:longlived}

In addition to the DM candidate, accidental symmetries can also give rise to other long-lived bound states. These can be $i)$ mesons made of light dark quark constituents, with masses below $\ldc$, and $ii)$ baryons and mesons with at least one heavy dark quark among their constituents and mass of order $\mH$. All these bound states can be metastable due to accidental species symmetries or G-parity, and must decay before $\sim 1\,$s  to avoid the constraints from Big Bang Nucleosynthesis (BBN) and indirect detection experiments. In this section we focus on long-lived light mesons, postponing the discussion of long-lived heavy hadrons to Sec.~\ref{subsec:degGUT}. Here we just notice, as already mentioned, that a split spectrum of dark quarks, i.e. $\mH \gg \mL$, is precisely needed to let all long-lived states decay before the onset of BBN. 

The decay of a metastable bound state is induced by the lowest dimension operator that breaks its accidental symmetry. In the case of light mesons, such an operator can be generated at any of the following scales:
\begin{itemize}
    \item the Planck scale $M_\text{P}$; 
    \item the GUT scale $M_{\text{GUT}}$ (by integrating out GUT gauge bosons and scalars);
    \item the mass scale $\mH$ (by integrating out dark quark GUT partners).
\end{itemize}
The lightest mesons in the dark matter sector are pseudo NGBs arising from the spontaneous breaking of its global symmetry. If charged under the SM gauge group, they have masses that are smaller than $\ldc$ by at most a factor $\sim 10$, see Eq.~(\ref{eq:mpiD}). For $\ldc \sim (50-100)\,$TeV, as required to reproduce the DM abundance with thermal relic dark baryons, this implies dark pion masses not much lighter than $(5-10)\,$TeV. Requiring that metastable dark pions have lifetime $\tau \lesssim 1\,$s, one obtains from Eq.~(\ref{eq:gamma}) the following upper bounds on the scale~$\luv$:
\begin{align}
  \label{eq:BS_D5}
      \text{For $D = 5$:} \quad \quad \luv &\lesssim 2 \times 10^{16} \,\mathrm{GeV} \times (k^2 \ndc)^{1/2} \bigg(\frac{m_{\pi_D}}{10\, \mathrm{TeV}} \bigg)^{3/2}\, , \\[0.1cm]
    \label{eq:BS_D6}
    \text{For $D = 6$:} \quad \quad \luv &\lesssim 1 \times 10^{10} \, \mathrm{GeV} \times (k^2 \ndc)^{1/4} \bigg(\frac{m_{\pi_D}}{10\, \mathrm{TeV}} \bigg)^{5/4} \, .
\end{align}
These bounds show that dark pions cannot be long-lived due to an accidental symmetry broken only by Planck-suppressed operators: this would imply a too long lifetime, in contrast with BBN bounds. The symmetry should arise as an accidental invariance of the Lagrangian below the GUT scale (if $D=5$) or the scale $\mH$, and be broken at the renormalizable level in the GUT theory. As discussed in greater detail in Sec.~\ref{sec:classification}, this excludes all GUT theories with accidental symmetries other than dark baryon number.

Among the light mesons, only $\bar V V$ can be long-lived because of an accidental G-parity. It was already mentioned that G-parity can be broken by the first two $D=5$ operators in Eq.~(\ref{eq:D5op}). In the following, we will thus focus on light mesons that are long-lived due to accidental species symmetries.

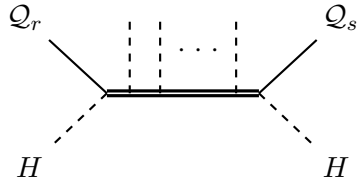
\begin{figure}
\centering
\begin{subfigure}[t]{0.4\textwidth}
\begin{tikzpicture}
\begin{feynman}
    \vertex(a0);
    \vertex[right=0.3cm of a0](ax1);
    \vertex[right=0.7cm of a0](ax2);
    \vertex[right=1.7cm of a0](ax3);
    \vertex[right=2cm of a0](a2);
    \vertex[right=0.28cm of ax2](d1);
    \vertex[right=0.5cm of ax2](d2);
    \vertex[right=0.72cm of ax2](d3);
    \vertex[above=1cm of ax1](aup1);
    \vertex[above=1cm of ax2](aup2);
    \vertex[above=1cm of ax3](aup3);
    \vertex[above left =1cm of a0](a1){\(\q_r\)};
    \vertex[below left =1cm of a0](a3){\(H\)};
    \vertex[above right =1cm of a2](a4){\(\q_s\)};
    \vertex[below right =1cm of a2](a5){\(H\)};
    \vertex[above=0.35cm of d1](dup1){\(\cdot\)};
    \vertex[above=0.35cm of d2](dup2){\(\cdot\)};
    \vertex[above=0.35cm of d3](dup3){\(\cdot\)};
    \diagram* {
        {
        (a0) --[double, very thick] 
        (a2),
        (a0) -- [thick]
        (a1),
         (a0) -- [thick, scalar]
        (a3),
        (a2) -- [thick]
        (a4),
        (a2) -- [thick, scalar]
        (a5),
        (ax1) -- [thick, scalar]
        (aup1),
        (ax2) -- [thick, scalar]
        (aup2),
        (ax3) -- [thick, scalar]
        (aup3),
       },
       };
\end{feynman}
\end{tikzpicture}
\end{subfigure}
\caption{Tree level Feynman diagrams that generate $D\geq 6$ species symmetry breaking operators at the scale $\mH$. A double line denotes the propagator of a dark quark GUT partner. The dashed lines coupled to the dark quark propagator denote either a Higgs field or a gauge field, where the latter reconstructs a covariant derivative.}
\label{fig:dim6_Htree}
\end{figure}

The bound of Eq.~(\ref{eq:BS_D6}) implies that operators with $D\geq 6$ must be generated necessarily at the scale $\mH\ll \mgut$ by integrating out dark quark GUT partners. The only renormalizable interactions that couple light dark quarks to their GUT partners are Yukawa interactions with the SM Higgs field $H$. Therefore, the $D\geq 6$ local operators that are generated at the scale $\mH$ and that can break accidental species symmetries are those obtained through the tree-level diagrams of Fig.~\ref{fig:dim6_Htree}.
For example, the $D=6$ operators are
\begin{equation}
  \bar\q_r \gamma_{\mu} \q_s H^{\dagger} \, i\!\overleftrightarrow{D}^{\mu} H \, , \qquad \quad \bar\q_r  \q_s H H H
  \, , \qquad \quad \bar\q_r i\!\!\not\!\! D \q_s HH \, ,
  \end{equation}
though the last one breaks the same global symmetries as the $5D$ operator $\bar\q_r \q_s HH$ and thus gives a subleading effect.~\footnote{Obviously, $D=6$ operators are of interest only when a $D=5$ operator breaking the same species symmetries cannot be written. This occurs for example when the $\bar\q_r \q_s$ bilinear forms an isosinglet with unit hypercharge.}
In general, operators with $D\geq 6$ are of the form $\bar\q_r (i\hspace{-0.12cm} \not\hspace{-0.16cm} D)^n H^{2+m}\q_s$ with $n+m\geq 1$.
All these operators are necessarily generated at the tree level since loop diagrams do not generate any new local operator that breaks accidental species numbers.~\footnote{In general, $D = 6$ operators can be generated at the scale $\mH$ also by diagrams with one insertion of some $D=5$ GUT-suppressed operator. This would lead to an effective scale $\luv = (\mgut \mH)^{1/2}$, which however does not satisfy the upper bound of Eq.~(\ref{eq:BS_D6}) for any $\mH \gtrsim \ldc$. This possibility can therefore be excluded.}
Notice that ``color species symmetries'', i.e. transformations under which dark species in the same $SU(3)$ irrep transform with the same phase, are not broken by the $D\geq 6$ operators, and must be therefore broken at the $D=5$ level if accidental. This is an obvious consequence of the fact that the Higgs field does not carry color, hence the $\bar\q_r \q_s$ bilinear is necessarily a color singlet. 

Clearly, the higher the dimension $D$, the lower the value of $\luv$ must be for the long-lived dark pion to decay before BBN. As a matter of fact, it is sufficient to consider operators with dimension up to 9.
Indeed, given the $SU(2)_L\otimes U(1)_Y$ quantum numbers of Tab.~\ref{tab:notation}, the $\bar\q_r \q_s$ bilinear has at most weak isospin 2 and hypercharge no bigger than~2. These quantum numbers can be matched by 4 Higgs fields, thus leading to $D=9$. Operators with $D>9$ do not break new global symmetries and can be thus neglected. For $D=9$ the bound on $\luv$ reads
\begin{equation}
\text{For $D = 9$:} \quad \quad \luv \lesssim 3 \times 10^{6} \,\mathrm{GeV} \times (k^2 \ndc)^{1/10} \bigg(\frac{m_{\pi_D}}{10\, \mathrm{TeV}} \bigg)^{11/10}\, .
\end{equation}

To conclude our discussion on the metastability of light mesons, we stress that models where species symmetries are not efficiently broken, so that some of the dark mesons decay after the onset of BBN, are to be discarded. This gives a selection criterion that will be used in the next section to identify realistic theories.

%
\section{Classification of Models}
\label{sec:classification}

In this section, we outline the general procedure used to classify candidate models. First, using a bottom-up approach, we identify the light dark quark sectors that include a DM candidate and that are free of dangerous long-lived states. The dark species content of the DM sector is selected by requiring that the dark color gauge group is in the confining phase and that SM gauge couplings remain perturbative in the UV. In a second step, for each DM model, we identify one or more \textit{minimal GUT parent theories}, i.e. the minimal sets of dark quark GUT partners that extend the original light field content into a full $SU(5)$ theory. 

The DM candidate should be the lightest dark baryon in the spectrum and have the correct quantum numbers under the SM.
As previously explained, we require that DM candidates be SM color singlets with zero hypercharge and integer weak isospin. Models that do not have light dark baryons with these quantum numbers can be immediately discarded since they are not viable. As an example, $\sundc$ models with $\q = T$ are not viable, since all (light) dark baryons have non-zero hypercharge.
If a dark baryon with the correct quantum numbers exists, then it can play the role of DM if it is the lightest in the spectrum. In order to establish if a theory is viable, one should then be able to identify the lightest baryon.
In general, this is a hard task, given the non-perturbative nature of the dark color dynamics.
However, as shown in Ref.~\cite{Antipin:2015xia}, one can obtain a possible answer under few a reasonable assumptions in the limit of small dark quark masses, $m \ll \ldc$. The argument goes as follows.

Dark baryons are dark color singlet states interpolated by operators with $\ndc$ dark quarks. Their mass receives contributions from the strong dynamics and from SM radiative corrections. One can thus parametrize light dark baryon masses as
\begin{equation} \label{eq:m_b}
m_B = a\, \ndc \, \ldc + b \, m + c \, \frac{\alpha_{SM}}{4 \pi} \, \ldc\, ,
\end{equation}
where $a,b,c$ are $\mathcal{O}(1)$ coefficients and we have used the large-$\ndc$ scaling to write the first term.
In the limit of zero quark masses and neglecting the weak gauging, the unbroken flavor symmetry in the confined phase of the dark color dynamics is $SU(N_{\text{DF}})$ for the case of $\sundc$ theories and $SO(N_{\text{DF}})$ for the case of $\sondc$ theories. Here $N_{\text{DF}}$ denotes the total number of dark quark fields in each case (see Eqs.~(\ref{eq:NDFforSUN}) and~(\ref{eq:NDFforSON})). The first term in Eq.~(\ref{eq:m_b}) will thus respect the flavor symmetry. In this limit, it is reasonable to assume that the lightest baryons have the smallest possible spin: spin $1/2$ for $\ndc$ odd, and spin 0 for $\ndc$ even.~\footnote{Theories with $N_{\text{DF}}=1$ are an exception, since in that case the lightest baryon has spin $\ndc/2$.} One can thus determine how the lightest baryons transform under the flavor symmetry by noticing that the baryon wave function must be fully symmetric in flavor and spin space. 
In particular, for $\sundc$ theories this means that the representations filled by the baryons under $SU(N_{\text{DF}})$ and $SU(2)$ spin
must have the same Young tableaux~\cite{Antipin:2015xia}. These are shown in Fig.~\ref{fig:spin_yt} for three values of $\ndc$.
\begin{figure}[t]
\centering
\text{$\ndc = 3: $}
\ytableausetup{centertableaux}
\yng(2,1)
\text{$\qquad\quad \ndc = 4: $}
\ytableausetup{centertableaux}
\yng(2,2)
\text{$\qquad\quad \ndc = 5: $}
\ytableausetup{centertableaux}
\yng(3,2)
\caption{Young tableaux of $SU(N)_{\text{DF}}$ corresponding to the smallest spin for $\ndc =3, 4, 5$.}\label{fig:spin_yt}
\end{figure}
In the case of $\sondc$ theories, one can start by considering the $SU(N_{\text{DF}})$ representation with the smallest spin and then decompose it into $SO(N_{\text{DF}})$ irreps. These will all have the same symmetry properties under the exchange of any two indices and will thus describe baryons with the smallest spin. Establishing which irrep is the lightest thus requires an additional assumption. Following Ref.~\cite{Antipin:2015xia}, we will assume that the smallest $SO(N_{\text{DF}})$ irrep is also the lightest. Direct inspection of the Young tableaux reveals that the smallest $SO(N_{\text{DF}})$ irrep is a fundamental if $\ndc$ is odd, a singlet if $\ndc = 4n$, with $n$ integer, and an antisymmetric if $\ndc = 4n +2$.

So far we have assumed vanishing dark quark masses and neglected SM radiative corrections. Both these effects, in general, will explicitly break the dark sector flavor symmetry down to the SM group.~\footnote{One could assume that quark masses have a common value, though SM radiative corrections will inevitably induce a splitting among different dark species. In our discussion, we will not make any assumption on the pattern of dark quark masses.} Correspondingly, the second and third terms in Eq.~(\ref{eq:m_b}) will split the masses of different SM fragments inside the previously identified lightest flavor representation. Which SM irrep becomes the lightest depends on the relative importance of dark quark masses and SM radiative effects. As long as $m \ll (\alpha_{SM}/4\pi) \ldc$, the radiative corrections (third term in Eq.~(\ref{eq:m_b})) dominate and the lightest baryon is the one with the smallest SM charge, i.e. the one transforming as the smallest SM irrep. For $(\alpha_{SM}/4\pi) \ldc  \lesssim m \ll \ldc$, on the other hand, the correction from dark quark masses (second term in Eq.~(\ref{eq:m_b})) dominates over radiative corrections and the lightest baryon is determined by its dark quark constituents. As an illustrative example, consider the $SU(3)_\text{DC}$ model with $\q = Q \+ \tilde D$. In this case, the dark color dynamics has an $SU(9)$ unbroken flavor symmetry in the limit of vanishing quark masses. The $SU(9)$ representation with the smallest spin is the 240, the corresponding Young tableau is the first of Fig.~\ref{fig:spin_yt}. If the dominant $SU(9)$ breaking is due to SM radiative corrections, then we expect that the lightest baryon is $QQ\tilde D$, transforming as a SM singlet: this is the smallest SM irrep contained in the 240 of $SU(9)$. If instead $(\alpha_{SM}/4\pi) \ldc  \ll m_Q, m_{\tilde D} \ll \ldc$ with $m_{\tilde D} < m_Q$, then the baryon $\tilde D\tilde D\tilde D$ with electromagnetic charge $-1$ is the lightest. In this case, the model is ruled out because it does not have a viable DM candidate. It is interesting to compare with two-flavor QCD: in that case EM radiative corrections are of order $1\,$MeV (in agreement with the naive estimate $\delta m_{EM} \sim (\alpha_{EM}/4\pi) \Lambda_{\text{QCD}}$ for $\Lambda_{\text{QCD}} \sim 1\,$GeV) and tend to make the neutron lighter than the proton. The contribution from quark masses turns out to be of the same order and slightly surpasses the radiative effect, giving $(m_n-m_p) \simeq 1.3\,$MeV. 

In the limit of large dark quark masses, $m \sim \ldc$, predictability is lost since the second term in Eq.~(\ref{eq:m_b}) becomes of the same order as the first term, and one cannot determine a priori which flavor irrep contains the lightest SM fragment.~\footnote{In principle, one might notice that the first term in Eq.~(\ref{eq:m_b}) scales as $\ndc$ in the large-$\ndc$ limit, and assume that it still dominates over the second for $m \sim \ldc$. We will not make use of this assumption in the following.}

In our classification, we select all models where a light dark baryon with the correct quantum numbers exists, without trying to establish that it is the lightest in the spectrum. This is a very conservative criterion, which excludes only those models that manifestly cannot give a DM candidate. In the majority of models obtained in this way, the argument given above shows that the lightest dark baryon is indeed a viable DM candidate for $m \ll \ldc$.
For the remaining models, we assume that the desired baryon can become the lightest in the spectrum  for an appropriate choice of dark quark masses of order $m \sim \ldc$, 

The list of models obtained using our procedure, together with their GUT parent theories, is reported in Appendix~\ref{appendix:models_list}. Compared to the analysis performed in Ref.~\cite{Antipin:2015xia}, we find more models as we relax some assumptions made in that study.~\footnote{In Ref.~\cite{Antipin:2015xia}, the selection of models with a viable DM candidate is performed by assuming a common value for the dark quark masses.  Furthermore, the so-called silver class models that were considered in~\cite{Antipin:2015xia} have at most two flavors of dark fermions; this leaves out many viable models.} Furthermore, while we find all the golden class models of Ref.~\cite{Antipin:2015xia}, some of their silver class models do not pass our selection criteria.

In the following, we describe our procedure in more detail by discussing separately the case of $\sundc$ and $\sondc$ theories.
In Sec.~\ref{sec:unification} we assess the quality of SM gauge coupling unification in each of the candidate models. In particular, we perform a further selection by keeping only models that improve the quality of unification obtained in the SM. For the selected models, we determine the allowed range of dark quark GUT partner masses.

\subsection{SU$(N)_{\text{DC}}$ Models}

We consider $\sundc$ theories with $\ndc \geq 3$. The case $\ndc=2$ is special, since the fundamental representation of $SU(2)$ is pseudo-real.  Theories with $SU(2)_{\text{DC}}$ dark color group are more similar to $\sondc$ ones, and can 
indeed be reformulated as $SO(3)_{\text{DC}}$ theories with fermions in the spinorial representation. We will not analyze them in this paper.

A DM sector is defined in terms of its light dark quark species content. In the case of $\sundc$ theories, each dark species is described by a Dirac spinor transforming as the fundamental of $\sundc$ and as an irrep $r_\text{SM}$ of the SM gauge group:
\begin{equation}
    \q_r  = (\square, \ r_{\text{SM}})\, , \qquad \text{$\q_r $ Dirac spinor field.}
\end{equation}
Only irreps $r_\text{SM}$ that are fragments of $SU(5)$ representations with rank equal or smaller than 2 are considered, as in Tab.~\ref{tab:notation}.
We adopt the notation of Ref.~\cite{Antipin:2015xia} and denote with $\tilde\q_r$ the dark species transforming in the conjugate SM irrep $\bar r_\text{SM}$ and as fundamentals of $\sundc$:
\begin{equation}
    \tilde \q_r  = (\square, \ \bar r_{\text{SM}})\, , \qquad \text{$\tilde \q_r$ Dirac spinor field.}
\end{equation}
Clearly, this alternative assignment of SM quantum numbers leads to new dark species in addition to those of Tab.~\ref{tab:notation} only if $r_{\text{SM}}$ is a complex representation.

We select candidate models by scanning over all possible light dark quark species contents and imposing the following requirements:
\begin{enumerate}
\item \textbf{Dark Colour Confinement}: The dark color gauge theory must be in the confining phase at low energy, this implies constraints on the number of dark colors  $\ndc$ and of light dark flavors $\ndf$ defined as
  \begin{equation}
    \label{eq:NDFforSUN}
\ndf = \sum_{r_\text{SM}} \dim(r_\text{SM})\, .  
\end{equation}
A necessary condition to have dark color confinement is that the 1-loop coefficient of the $\sundc$ $\beta$-function be negative:~\footnote{We use the renormalization group equation $d \alpha_i^{-1}/d \log Q = - \beta_i/ 2 \pi$ to define gauge $\beta$-function coefficients $\beta_i$.}
\begin{equation}
\label{eq:beta_dc_sun}
\beta_{DC} = -\frac{11}{3} \ndc + \frac{2}{3}  \ndf < 0 \, .
\end{equation}
This, however, is not a sufficient condition for confinement. For the dark color gauge group to confine, the number of light dark flavors $\ndf$ in the theory should be below the lower edge of the conformal window: $\ndf < N_{\text{conf}}$. The value of $N_{\text{conf}}$ cannot be derived in perturbation theory and has been determined for QCD to be equal to $N_{\text{conf}} = 12$ by lattice simulations (see~\cite{DeGrand:2015zxa}). Following~\cite{Appelquist:1996dq}, we generalize such a result for an arbitrary number of dark colors $\ndc$ and require that DM sectors have:
\begin{equation}
  \ndf \leq 4 \ndc \, .
\end{equation}
\item \textbf{Perturbativity of SM couplings}: Unification of SM gauge couplings can occur in our theories if these remain perturbative when extrapolated to energies of the order of the unification scale. As a necessary condition for this to happen, we require that after including the contribution of light dark quarks to their $\beta$-functions, the SM gauge couplings do not develop Landau poles below the scale $M_{GUT}^{SM} = 5.3 \times 10^{14}$ GeV. This is not a sufficient condition because the full RG evolution includes the contribution of dark quark GUT partners as well. The absence of Landau poles can be recast as the following upper bounds on the $\beta$-function coefficients:
\begin{equation}
 \Delta \beta_Y \lesssim 18, \quad \quad \quad \Delta \beta_2 \lesssim 12, \quad \quad \quad \Delta \beta_3 \lesssim 11 \, ,
\end{equation}
which in turn translates into upper bounds on the number of dark colors $\ndc$ and light dark flavors $\ndf$.
 \item \textbf{Presence of DM candidate}: We select DM models with at least one potential DM candidate, i.e. a dark baryon with no color, no EM charge, and no hypercharge. For example, for $\ndc = 3$ the DM candidates are given by:
\begin{equation}
    QQ\tilde{D}, ~ DDU
  \end{equation}
for theories with colored dark fermions, and 
\begin{equation}
    LLE, ~ LLT, ~ VE\tilde{E}, ~ VL\tilde{L}, ~ NE\tilde{E}, ~ NL\tilde{L},~ VVV,~ VVN, ~VNN,~ NNN.~\footnote{The DM candidate $\T{T} T N$ is excluded since the model with dark quarks $\T{T} \+ T \+ N$ is forbidden by the condition on $SU(2)_L$ and $U(1)_Y$ perturbativity.} 
\end{equation}
for theories where dark fermions have only EW charges. Whether a potential candidate is the lightest in the spectrum and thus plays the role of DM depends on the value of the dark quark masses. For example, consider the $V\+ N$ model with $\ndc=3$. In this case, spin 1/2 baryons transform in the first Young tableau of Fig.~\ref{fig:spin_yt}. Hence, baryons $VVN, \, VVV$ and $VNN$ have spin 1/2, while $NNN$ has spin 3/2.
In the limit of small quark masses, $m_{V, N} \ll \ldc$, the lightest baryon is expected to be $VVN$, since it transforms as an SM singlet, while $VNN$ and $VVV$ are (atleast) EW triplets. However, by taking $\ldc (\alpha_{SM}/4\pi) \ll m_{V, N} \ll \ldc$ with $m_N < m_V$, the contribution to the baryon mass due to dark quark masses can overcome radiative corrections, and $NNV$ can become the lightest baryon. Furthermore, in the limit of large dark quark masses, the spin 3/2 baryon $NNN$ may become the lightest in the spectrum, although establishing this requires a non-perturbative control over the dark color dynamics.

As explained previously, not all the selected models have a DM candidate in the limit of small masses. For example, consider the $T\+ E\+ N$ model with $\ndc=3$. In this case, none of the spin 1/2 baryons is a good DM candidate. We do not discard the model, however, since it can be viable for large dark quark masses, $m_{T, E} > m_N \sim \ldc$, as long as the spin 3/2 baryon $NNN$ is the lightest in the spectrum.

\item \textbf{Absence of dangerous long-lived light hadrons:} We select DM models where G-parity and all species symmetries are broken either by Yukawa couplings or by the allowed higher-dimensional operators discussed in Sec.~\ref{subsec:longlived}. The full list of operators for the $\sundc$ case is shown in Table~\ref{table:SU(N)operators}.
%
\begin{table}[t]
 \begin{center}
\begin{tabular}{ c c c  }
\textbf{operator} & \textbf{type} & \textbf{dim} \\[0.1cm]
  \hline
  & & \\[-0.35cm]
  $\bar{Q}\tilde{D}H$, $\bar{Q}\tilde{U}H^c$, $\bar{L}\tilde{E}H$, $\bar{L}\tilde{T}H$, $\bar NLH$, $\bar VLH$ & $\bar\q_r\q_s H$ & 4 \\[0.15 cm]
 $\bar L \sigma^i\tilde L H^{\dagger} \sigma^i H^c$, $\bar NV^iH^\dagger\sigma^i H$, $\bar EV^iH^{c \dagger}\sigma^i H$,
  $\bar ET^iH^{\dagger}\sigma^i H$ &  $\bar{\q}^r\q_s HH$ & 5 \\[0.05cm]
  $\bar T^i N H^{c \dagger}\sigma^i H$, $\epsilon^{ijk} \bar V^i V^j H^\dagger \sigma^k H$, $\epsilon^{ijk} \bar T^i V^j H^{c \dagger} \sigma^k H$ & & \\[0.15cm]
  $\bar N\sigma^{\mu\nu}V^iW_{\mu\nu}^i$, $\bar E\sigma^{\mu\nu}T^iW_{\mu\nu}^i$, $\bar G^a\sigma^{\mu\nu}NG_{\mu\nu}^a$
 & $\bar \q_r \sigma^{\mu\nu} \q_s F_{\mu\nu}$ & 5 \\[0.05cm]
  $\bar S\sigma^{\mu\nu}UG_{\mu\nu}$, $\bar V \sigma^{\mu\nu} V B_{\mu\nu}$ & &  \\[0.15cm]
$\bar LE HHH$,  $\bar LT HHH$, $(\bar{T}^i H^{c \dagger} \sigma^i H) (\tilde{L} H^{\dagger})$,$(\bar{V}^i H^{c \dagger} \sigma^i H) (L H^{\dagger})$ & $\bar\q_r\q_s HHH$ & 6 \\[0.15cm]
 $\bar D\gamma^\mu U H^{c \dagger}i\!\overleftrightarrow{D}_{\!\!\mu} H$, $\bar E\gamma^\mu N H^{c \dagger}i\!\overleftrightarrow{D}_{\!\!\mu} H$, $\bar{Q} \gamma^{\mu} \tilde{X} H^{c \dagger}i\!\overleftrightarrow{D}_{\!\!\mu} H$ & $\bar\q_r\gamma^\mu\q_s H iD_\mu H$ & 6 \\[0.15cm]
$\bar{V}^i \gamma^{\mu} \tilde{T}^i H^{c \dagger}i\!\overleftrightarrow{D}_{\!\!\mu} H$,  $\bar{\tilde{L}} \gamma^{\mu} L H^{c \dagger}i\!\overleftrightarrow{D}_{\!\!\mu} H$, & & \\[0.15cm]
 $\bar{X} \gamma^{\mu} U H (H^{c \dagger}i\!\overleftrightarrow{D}_{\!\!\mu} H)$, $\bar{T}^i \gamma^{\mu} L \sigma^i H (H^{c \dagger}i\!\overleftrightarrow{D}_{\!\!\mu} H)$ & $\bar\q_r\gamma^\mu\q_s H H iD_\mu H$ & 7 \\[0.15cm]
$ \bar{E} L H  (H^{c \dagger}i\!\overleftrightarrow{D}_{\!\!\mu} H)$ & &  \\[0.15cm]
  $\bar T^i\tilde{T}^j (H^{c \dagger}\sigma^iH)(H^{c \dagger}\sigma^jH)$, $\bar{T}^i V^j (H^{c \dagger} \sigma^i H)(H^{\dagger} \sigma^j H)$  & $\bar\q_r\q_s HHHH$ & 7 \\[0.15cm]
  $ \bar{T}^i \gamma^{\mu} \tilde{E} (H^{c \dagger} \sigma^i H) (H^{c \dagger}i\!\overleftrightarrow{D}_{\!\!\mu} H)$ & $\bar\q_r\gamma^\mu\q_s H H  H iD_\mu H$  & 8 \\[0.15cm]
  
 $\bar E\tilde{E} (H^{c \dagger}D_\mu H)^2$, $\bar E \tilde T^i (H^{c \dagger}\sigma^i D_\mu H)(H^{c \dagger}D^\mu H)$, $\bar{T} \tilde{T} (H^{c \dagger}D_\mu H)^2$  & $\quad \bar\q_r\q_s (H D_\mu H)^2$ & 9
\end{tabular}
\end{center}
\caption{List of lowest-dimensional operators that break G-parity and species symmetries in $\sun$ theories. Each term stands for two operators, corresponding to the two possible choices of chiralities of the fermionic fields. When more operators exist with the same quantum numbers, we show only one with the lowest dimensionality.}
\label{table:SU(N)operators}
\end{table}
As already explained in Sec.~\ref{subsec:longlived}, no operator couples an $SU(3)$ colored dark fermion to one with purely SM electroweak quantum numbers. In other words, the operators of Table~\ref{table:SU(N)operators} preserve color species symmetries. Regarding $SU(3)$-charged dark fermions, we note that the allowed operators mix only $\T{Q},$ $D$ and $U$ (or  $Q,$ $\T{D}$ and $\T{U}$).
This means that any viable DM model must be either a combination of $SU(3)$ singlet dark fermions or a combination of $\T{Q},$ $D$ and $U$ (or  $Q,$ $\T{D}$ and $\T{U}$).
\end{enumerate}

For each selected DM model, we identify its minimal GUT parent theories. These are defined as the minimal sets of $SU(5)$ GUT representations 
that contain both the light dark fermions and the GUT partners needed to generate the required symmetry-breaking higher-dimensional operators.
As discussed in Sec.~\ref{subsec:longlived}, the latter must arise below the Planck scale by integrating out states with mass much higher than $\ldc$. This implies, by our definition of minimal parent theory, that the GUT Lagrangian must have no accidental symmetry. This is a strong requirement on the allowed GUT theories. Notice that a given DM model can have more than one GUT parent theory, while the same GUT parent theory can lead to several low-energy DM models. 

To illustrate how we identify parent GUT theories, consider the following two examples of $SU(3)_{DC}$ theories:
 \begin{itemize}
 \item $Q\oplus \tilde{D}$ model: In this case the DM candidate is $QQ\tilde{D}$, and the species symmetry corresponding to the relative phase between $Q$ and $\tilde D$ is broken by the allowed Yukawa operator $\bar{Q}\tilde{D}H$. There exist two minimal GUT parent theories: the first, given by $5 \+ 10$, has the Yukawa interaction $\bar{\Psi}_5 \phi_5^{\dagger} \Psi_{10}$; the second, given by $5\+ 15$, has the Yukawa term $\bar{\Psi}_5 \phi_5^{\dagger} \Psi_{15}$.
Here, $\Psi_5$, $\Psi_{10}$ and $\Psi_{15}$ are Dirac fields transforming respectively as $5$, $10$ and $15$ of $SU(5)$ ($\Psi_5$ contains $\tilde{D}$, while $\Psi_{10}$ and $\Psi_{15}$ contain $Q$), while $\phi_5$ is the GUT scalar field, transforming as a $5$, that contains the SM Higgs field.
 \item $L  \+ E$  model: In this case, the DM candidate is $LLE$, and the species symmetry corresponding to the relative phase between $L$ and $E$ is accidental in the low-energy theory (no Yukawa interaction is allowed between $L$ and $E$). Unlike the $Q + \T{D}$ model, the simplest GUT embedding of $L$ and $E$ fragments into $\bar{5} \+ 10$ cannot be considered as a viable parent GUT theory. This is because this GUT theory itself has an accidental species number symmetry (no Yukawa interaction between $\bar 5$ and $10$ can be written), which is inherited by the low-energy DM sector. In other words, a $\bar{5} \+ 10$ theory fails to generate the required symmetry-breaking higher-dimensional operators. One must then introduce additional \textit{auxiliary} GUT multiplets to have a sufficiently sizable violation of all the species symmetries. There exist two minimal GUT parent theories that do so: $\bar{5} \+  10  \+  5  \+  1$ and $\bar{5} \+ 10 \+ 5 \+ 24$, with auxiliary multiplets $5, 1$ and $5,24$ respectively. All species symmetries are broken by Yukawa operators, for example in the first parent theory one can write $\bar{\Psi}_{10}    \phi_5 \Psi_5$,  $\bar{\Psi}_{1}    \phi_5 \Psi_{\bar{5}}$ and $\bar{\Psi}_{5}    \phi_5 \Psi_{1}$. By integrating out the auxiliary fields one generates the $D=6$ operator $\bar{L}EHHH$, which breaks the species symmetry of the DM theory. Notice that the same GUT parent theories of the $L \+ E$ model also give rise to the $L \+ E \+ \T{L} \+ N$ DM model.
\end{itemize}

\subsection{SO$(N)_{\text{DC}}$ Models}\label{sec:models_so_n}

In the case of $\sondc$ theories, dark species are described by Dirac spinors or Majorana spinors depending on whether they transform as complex or (pseudo)real irreps $r_\text{SM}$ of the SM gauge group:
\begin{equation}
\q_r  = (\square, \ r_{\text{SM}})\, , \qquad \text{$\q_r $ Dirac (Majorana) spinor field if $r_\text{SM}$ complex (real).}
\end{equation}
Since the fundamental representation of $\sondc$ is real, there is no distinction in this case between $\q$ and $\tilde\q$. Other two important consequences are: (i) there is no distinction between baryons and antibaryons; (ii) dark color singlet operators that excite dark hadrons can be constructed with both $\q$ and $\q^c$ operators, where $\q^c \equiv i\gamma^2 \q^*$. For example, dark mesons can be interpolated by both $\bar\q_r \q_s$ and $\bar \q^{c}_r \q_s$ operators. 
Both species numbers associated to $\q_r$ and $\q_s$ must therefore be broken to let both types of dark mesons decay.

Similarly to the case of $SU(N)_{\text{DC}}$ theories, we identify all the possible species symmetry breaking operators including the allowed non-renormalizable interactions. The full list of operators can be obtained from Table~\ref{table:SU(N)operators} by simply replacing $\tilde{\q}$ with $\q^c$.

Candidate DM models are selected by imposing the same criteria as for $\sundc$ theories: dark color confinement, perturbativity of SM gauge couplings, presence of a DM candidate, and absence of dangerous long-lived states. For the perturbativity criterion, we use the $\sondc$ one-loop $\beta$-function
\begin{equation}
        \beta_{DC}=-\frac{11}{3}(\ndc-2)+\frac{2}{3}N_{\text{DF}}\, ,
\end{equation}
where $\sondc$ generators in the fundamental representation are normalized as Tr$(T^a T^b)=\delta^{ab}$, and the number of dark flavors is defined by
\begin{equation}
  \label{eq:NDFforSON}
\ndf = \sum_{\substack{\text{complex} \\ r_\text{SM}}} 2\dim(r_\text{SM}) + \sum_{\substack{\text{real} \\ r_\text{SM}}} \dim(r_\text{SM}) \, .  
\end{equation}
We require $\beta_{DC}<0$ but do not impose the stronger bound $\ndf < N_{\text{conf}}$, since we are not aware of any lattice study evaluating $N_{\text{conf}}$ for $SO(N)$ theories. 

As for $\sundc$ theories, in most models, though not in all of them, the lightest dark baryon is expected to be a good DM candidate in the limit of small dark quark masses. For example, according to the argument presented earlier in this section, for $\ndc=4$ the lightest baryon is expected to transform as a singlet of $SO(N_{DF})$. In the model with $\q = T$, this leads to the SM singlet DM candidate $TT^cTT^c$.
For $\ndc =5$, instead, the lightest baryons are expected to fill a fundamental of $SO(N_{DF})$. In the $V\+ L$ model, this includes $LL^cLL^cV$, which is a good DM candidate and transforms as an EW triplet. In the $E\+ L$ model, on the other hand, the fundamental of the $SO(6)$ flavor group does not contain any viable DM candidate. The model can still be realistic if, for large dark quark masses $m_{L, E}\sim \ldc$, the lightest baryon is $EE^cLLE$ or $LL^cLLE$.

\section{Gauge Coupling Unification}
\label{sec:unification}

In the previous section, we selected candidate dark matter models and identified their GUT parent theories. Such GUT parent theories are further screened in this section based on the quality of their gauge coupling unification. 
The basic requirement we impose is very mild, as we want to cast a net wide enough not to lose any potentially interesting theory: we select only theories that equal or improve the quality of unification in the SM. This is a rather minimal requirement (SM unification has already been ruled out by the current lower bound on the proton lifetime~\cite{Super-Kamiokande:2016exg}), which leaves open the possibility that (large) threshold corrections exist at the GUT scale. Yet, only few theories turn out to satisfy it. For those that do, we impose a final condition: metastable heavy hadrons (with masses of order $\mH$) must decay before BBN. This sets a lower limit on the GUT partner mass scale.

In the following, we discuss our selection and report the results for both $\sundc$ and $\sondc$ theories. We start by assuming that all dark quark GUT partners in a given theory have a common mass scale $\mH$. This simplifying assumption is then relaxed in the case of a few benchmark theories.

\subsection{Unification Criteria}
\label{sec:unificationcriteria}

The evolution of SM gauge couplings is determined by means of the following simplified 1-loop RG equations:
 \begin{equation}
    \label{eq:running}
    \frac{1}{\alpha_i(\mu)} = \frac{1}{\alpha_{i}(m_Z)} - \frac{\beta_{i}^{SM}}{2 \pi} \log \frac{\mu}{m_Z} -  \frac{\beta_{i}^{L}}{2 \pi} \log \frac{\mH}{\Lambda_{DC}} - \frac{\beta^{SU(5)}}{2 \pi} \log \frac{\mu}{\mH} + \Delta_i\, ,
\end{equation}
where $\beta_{i}^{SM} = (41/10, \, -19/6, \, -7)$ are the 1-loop SM $\beta$-functions (we adopt the standard definition $g_1 = \sqrt{5/3}g_Y$, where $g_Y$ is the SM hypercharge coupling). The contribution of light dark quarks is encoded by $\beta_i^{L}$ and turns on at $\ldc$, while that of their GUT partners is encoded by $(\beta^{SU(5)}-\beta_i^{L})$ and turns on at $\mH$ (when we consider a split spectrum of dark quark GUT partners, the RG equations are modified accordingly). For simplicity, we do not consider any contribution from the strong dynamics below the dark confinement scale. The parameters $\Delta_i$ encapsulate all the IR and UV threshold corrections and are set to zero in our calculation. As explained above, we assume that the UV threshold corrections at the GUT scale account for an imperfect unification.

\begin{figure}[t!] 
\begin{subfigure}{0.5\textwidth}
\includegraphics[width=\linewidth]{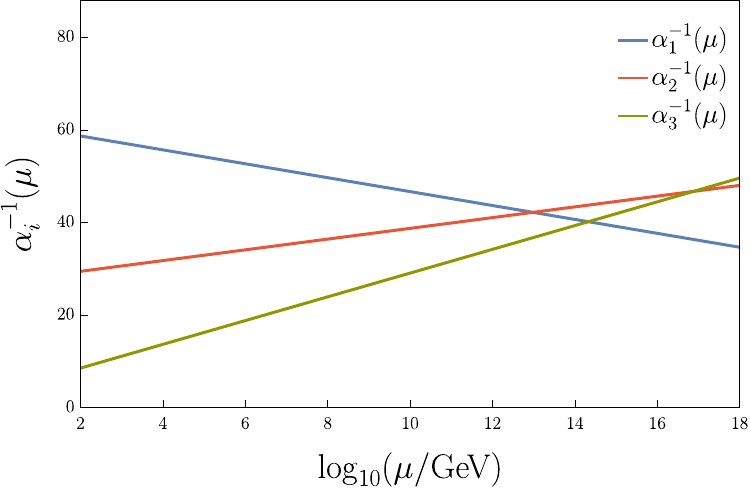}
\end{subfigure}\hspace*{\fill}
\begin{subfigure}{0.5\textwidth}
\includegraphics[width=\linewidth]{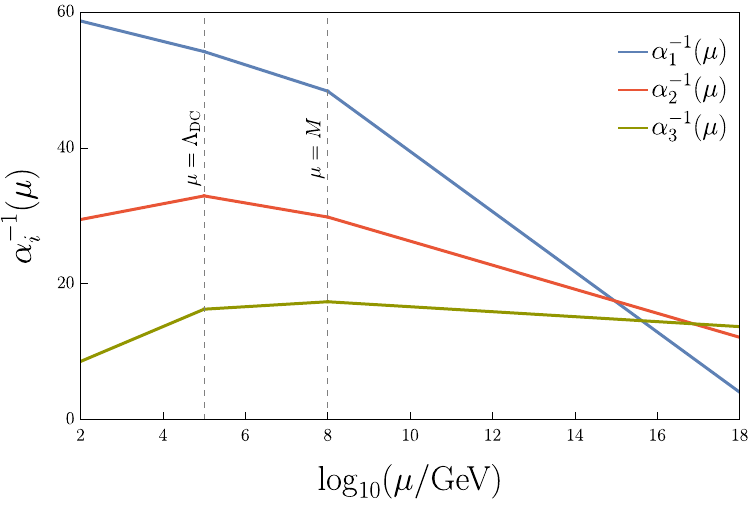}
\end{subfigure}
\caption{One-loop renormalization group evolution of the SM gauge couplings in the SM \textit{(left)} and in the $SU(3)_{DC}$ model $Q\+ \tilde D$ with $\mH=10^8\,$GeV \textit{(right)}.}
\label{fig:delta_cartoon}
\end{figure}

When evolved to high energy using Eq.~(\ref{eq:running}), the SM gauge couplings approach each other and form a triangle in the plane ($\log_{10} \mu, \alpha_{i}^{-1}(\mu))$, see Fig.~\ref{fig:delta_cartoon}. 
The quality of unification is quantified based on the size and the location of the triangle. In particular, we define the unification scale $\mgut$ and the unified coupling $\alpha_{GUT}(\mgut)$ as respectively the horizontal position and the inverse of the vertical position of the barycenter of the triangle.
When applied to the SM, these definitions give a unification scale equal to $\mgut^{\text{SM}}$ $=$ $5.3 \times 10^{14}$ GeV, which is too small to be compatible with the bounds on proton decay set by Super-Kamiokande~\cite{Super-Kamiokande:2016exg}. 

We scan theories by requiring that they equal or improve the quality of gauge coupling unification of the SM. In practice, for each given theory we vary the GUT partner mass scale $\mH$ and impose the following constraints:
\begin{itemize}
\item \textbf{Precision Constraint:} the area of the triangle formed in the plane $(\log_{10} \mu, \alpha_{i}^{-1}(\mu))$ must be smaller than or equal to the one obtained in the SM: \ \  $\mathcal{A}_\Delta \leq \mathcal{A}_\Delta^{SM}$.  
\item \textbf{Perturbativity Constraint:} the SM and dark color gauge couplings must remain perturbative up to the unification scale: \ \ 
  $\alpha_{GUT}(\mgut) < 4 \pi$,\ $\alpha_{DC}(\mgut) < 4 \pi$.
\item \textbf{Constraint on the Unification Scale:} the GUT scale must be larger than the SM unification scale and smaller than the Planck scale: \ \   $5.3 \times 10^{14}\,\text{GeV}\leq \mgut \leq M_P$.
\end{itemize}
Theories that cannot satisfy the above requirements for any value of the GUT partner mass in the range $\ldc < \mH \leq \mgut$ are discarded. As already mentioned, we perform our main selection by assuming a single mass for all the dark quark GUT partners. This simplifying assumption is then relaxed for a few theories to assess its impact. The results of our analysis are reported below, respectively for the case of degenerate and split mass spectrum of dark quark GUT partners.

\subsection{Viable Theories: degenerate GUT partner spectrum}
\label{subsec:degGUT}

We find that none of the $\sondc$ theories selected in the previous section can satisfy the unification criteria under the assumption of a common mass scale for the dark quark GUT partners.
Among $\sundc$ theories, only the following ones do:~\footnote{The authors of Ref.~\cite{Antipin:2015xia} found three theories which lead to good gauge coupling unification. The first one, the $SU(3)_{DC}$ model $Q+\tilde D$ with $5\+ 10$ parent theory, is also in our list. The other two theories do not pass our selection, in particular: the $SO(3)_{DC}$ model $V$ with $24$ parent theory has a too low unification scale, and the $SU(3)_{DC}$ model $Q\+ D\+ U\+ L$ does not pass the selection of Sec.~\ref{sec:classification}.}
\begin{itemize}
\item $\mathbf{Q \+ \T{D}}$ with $\mathbf{5\oplus 10}$ parent theory and $\ndc = 3$.

Our unification requirements are satisfied for masses of GUT partners ($U, E$, and $\tilde L$) in the range $1  \times 10^7\,$GeV $< \mH < 9 \times 10^{14}\,$GeV, corresponding to a unification scale $3 \times 10^{15}\,$GeV $ < \mgut< M_P$. The upper and lower bounds on $\mH$ come from imposing the constraint on $\mgut$.
As already noticed in Ref.~\cite{Antipin:2015xia}, this theory achieves perfect (tree-level) gauge coupling unification at a scale  $\mgut =1 \times 10^{17}\,$GeV for $\mH = 1 \times 10^{11}$ GeV.
\item $\mathbf{Q \+ \T{D}}$ with $\mathbf{5\oplus 15}$ parent theory and $\ndc = 3$.

Our unification requirements are satisfied for masses of GUT partners ($T, S$ and $\tilde L$) in the range $4 \times 10^{11}\,$GeV $< \mH < 9 \times 10^{14}\,$GeV, corresponding to a unification scale $1 \times 10^{17}\,$GeV $ < \mgut < M_P$. The lower bound on $\mH$ comes in this case from imposing the perturbativity constraint. It follows that this theory cannot achieve perfect unification in a perturbative regime.
\item $\mathbf{N}$ with $\mathbf{24}$ parent theory.

This theory has the same differential running as the SM, hence it always predicts $M_{GUT} = M_{GUT}^{SM}$ and $\mathcal{A}_\Delta = \mathcal{A}_\Delta^{SM}$. The perturbativity requirement can be always satisfied, for any value of $\ndc$, as long as the mass of the GUT partners ($ V, X$ and $G$) is sufficiently large. For example, one must require $\mH > 1 \times 10^9\,$GeV for $\ndc = 3$.
  On the other hand, even for $\mH\sim\mgut$, too large values of $\ndc$ drive the $SU(5)$ gauge coupling strong above $\mgut$. In particular, we find that for $\mH=\mgut$ the $SU(5)$ gauge coupling has a Landau pole below $M_P$ for $\ndc > 7$. 
\end{itemize}

\begin{figure}[t!] 
\centering
\includegraphics[width=0.6\linewidth]{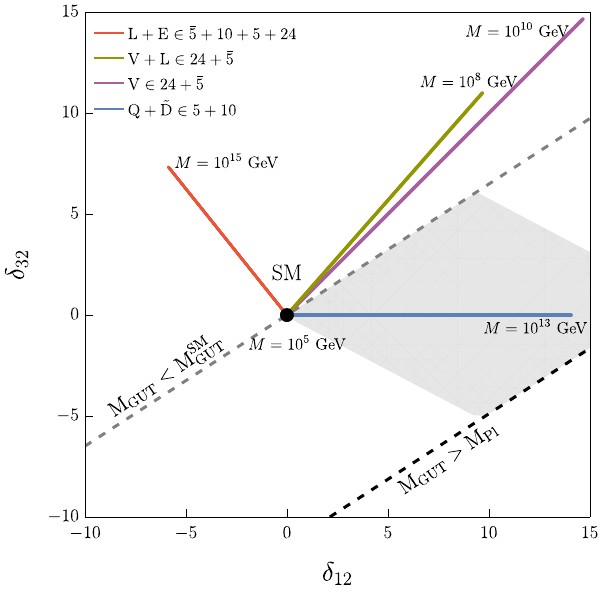}
\caption{\small{Trajectories in the $(\delta_{12}, \delta_{32})$ plane traced by varying the GUT partner mass $\mH$ in a few benchmark $\sundc$ theories, with $\ndc =3$. Theories in the gray-shaded region satisfy the unification criteria and thus lead to a better unification than in the SM. The black point marks the SM prediction $\delta_{12} = 0, \ \delta_{32} = 0$.
}}
\label{fig:allowedmodels}
\end{figure}

The other theories are discarded since they all predict a unification scale smaller than the SM one. This happens because, with the exception of $Q\+ \tilde D\+ \tilde U$ and $\tilde D\+ \tilde U$ models, their light dark quarks have electroweak quantum numbers but are color singlets. It turns out that increasing the value of the unification scale compared to the SM one requires new colored matter (see for example,~\cite {Giudice:2004tc}).
This conclusion is visually summarized by Fig. \ref{fig:allowedmodels}, which shows the trajectories traced in the plane $(\delta_{12}, \delta_{32})$ by varying the GUT partner mass $\mH$ in a few benchmark theories with $\ndc=3$.
Here $\delta_{ij}$ is defined to encode the contribution to the differential running from light dark fermions:
\begin{equation}
  \delta_{ij} = -  \frac{\beta_{i}^{L}- \beta_{j}^{L}}{2 \pi} \log \frac{\mH}{\ldc} \, .
\end{equation}
The grey region in Fig.~\ref{fig:allowedmodels} corresponds to values of $\delta_{12}$ and $\delta_{32}$ satisfying our unification criteria.
Only the $Q\+ \tilde D$ model enters this region.

A high-quality unification requires dark quark GUT partners with masses much larger than $\ldc$. In other words, unification implies a split spectrum of dark quarks. There is however another reason why GUT partners must be much heavier than light dark quarks: some of the hadrons they form are metastable and must decay earlier than $\sim 1\,$s to avoid the strong constraints set by BBN. The metastability of heavy hadrons is a consequence of accidental species numbers acting on the dark quark GUT partners.

Consider for example the $Q\+ \tilde D$ model with $5\+ 10$ GUT parent theory~\footnote{For more details, see~\cite{BCV}.}. There are three GUT partners, $\tilde L$, $E$, and $U$, and only one Yukawa coupling involving them: $\bar E H \tilde L + h.c.$. This implies 2 accidental species symmetries, which are broken by $D=6$ operators of the kind $\bar Q \tilde L q \ell$, $\bar U \tilde D u^c d^c$, $\bar Q \tilde L \bar{\tilde D} U$, generated at the GUT scale. For $\mH \ll \mgut$, the lightest hadrons charged under the new species symmetries will be long-lived, since their decay is induced only by the symmetry-breaking higher-dimensional operators. Using the estimate of Eq.~(\ref{eq:gamma}) with $\luv = \mgut$, $k\sim 1$ and $D=6$, one obtains a lifetime shorter than $1\,$s for GUT partner masses of order $\mH \gtrsim 10^9 \,$GeV. Similar considerations hold also for the other two theories that pass the unification criteria. In the $Q\+ \tilde D$ model with $5\+ 15$ parent theory there are three GUT partners ($\tilde L$, $T$ and $S$) and one Yukawa coupling involving them: $\bar T^i H \sigma^i \tilde L+h.c.$. This implies, as in the case of the $5\+ 10$ theory, two accidental species symmetries, broken by $D=6$ operators. The $N$ model with $24$ parent theory has four GUT partners ($G$, $V$, $X$, and $\tilde X$) and no Yukawa coupling involving them. There are thus four species symmetries, broken by $D=5$ and $D=6$ operators.

We thus conclude that, in all the cases, GUT partners must be much heavier than $\ldc$ to give a successful unification and to avoid the BBN bounds from the decay of metastable heavy hadrons. In Sec.~\ref{sec:mass_split} we will see that such a split spectrum involves some amount of fine tuning.

\subsection{Viable Theories: non-degenerate GUT partner spectrum}
\label{section:nondegmass}

\begin{figure}[t!] 
\begin{subfigure}{0.5\textwidth}
\includegraphics[width=\linewidth]{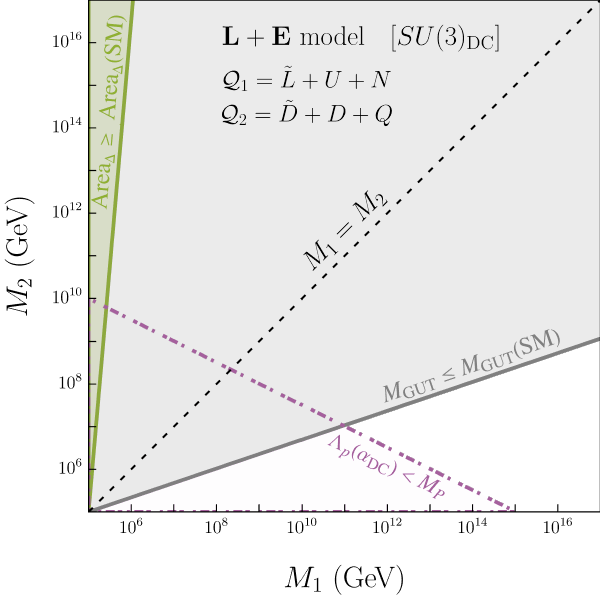}
\end{subfigure}\hspace*{\fill}
\begin{subfigure}{0.5\textwidth}
\includegraphics[width=\linewidth] {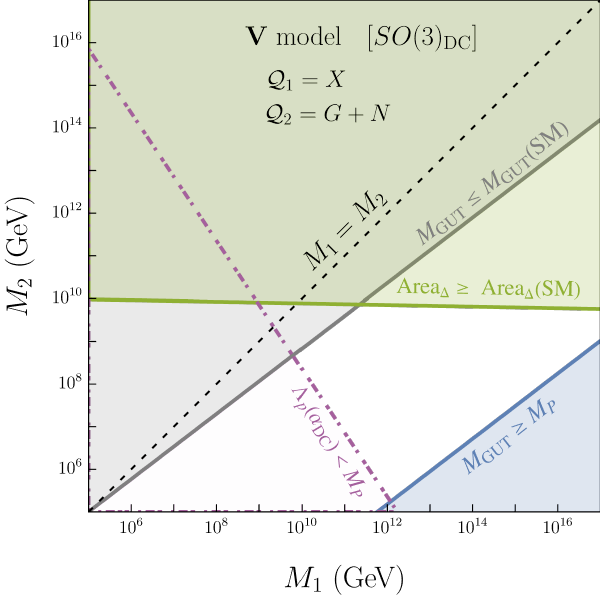}
\end{subfigure}
\caption{\small{Quality of gauge coupling unification for a spectrum of dark quark GUT partners characterized by two mass scales $\mH_1$ and $\mH_2$. \textit{Left plot:} $SU(3)_{DC}$ model $L \+ E$ with $\bar{5} \+ 10 \+ 5 \+ 1$ parent theory; GUT partners $\tilde L, U, N$ and $\tilde D, D, Q$ have mass equal to  $\mH_1$ and $\mH_2$ respectively. \textit{Right plot:} $SO(3)_{DC}$ model $V$ with $24$ parent theory; GUT partners $X$ and $G, N$ have mass equal to $\mH_1$ and $\mH_2$ respectively. Points in the white region satisfy all the unification criteria. The green region is excluded by the precision constraint $\mathcal{A}_\Delta \leq \mathcal{A}_\Delta^{SM}$. The gray and blue regions are excluded because they lead respectively to $\mgut < \mgut^{SM} = 5.3 \times 10^{14}\,$GeV and $\mgut > M_P$. Below the dot-dashed purple line, the dark color gauge coupling develops a Landau pole below the Planck scale.}}
\label{fig:nondeg}
\end{figure}

We noticed that obtaining a large enough GUT scale requires the contribution of colored fermions to the differential running (see also Ref.~\cite{Giudice:2004tc}). This suggests that more theories will pass the unification requirements if we allow for a split spectrum of dark quark GUT partners. To verify this, we re-analyzed two of the excluded theories, the $SU(3)_{DC}$ model $L \+ E$ with $\bar{5} \+ 10 \+ 5 \+ 1$ parent theory and the $SO(3)_{DC}$ model $V$ with $24$ parent theory, and assumed that the spectrum of dark quark GUT partners is characterized by two different mass scales, denoted by $\mH_1$ and $\mH_2$. We find that in both cases there exist suitable mass splittings for which the unification criteria are all fulfilled. For example, Fig.~\ref{fig:nondeg} shows the allowed region in the plane $(\mH_1,\mH_2)$ for the following mass splittings: $\q_1 = \T{L}\+ U \+ N$ with mass $\mH_1$ and $\q_2 = D \+ \T{D} \+ Q$  with mass $\mH_2$ in the $\bar{5} \+ 10 \+ 5 \+ 1$ theory; $\q_1 = X,$ with mass $\mH_1$ and $\q_2 = G\+ N$ with mass $\mH_2$ in the $24$ theory.
We believe that the same conclusion can be generalized to other theories among those selected in the previous section: the unification requirements can be in general satisfied for suitably split spectra of GUT partners. Although this enlarges in principle the list of theories fulfilling our requirements, split spectra of dark quark masses are obtained only at the price of some fine tuning. The more mass scales a theory has, the more tuned it is. This is discussed in detail in the next section.

 \section{Mass Splitting}
\label{sec:mass_split}

We have seen that a hierarchy between the mass of the light dark quarks (those belonging to the DM sector) and that of their GUT partners is required to let metastable heavy hadrons decay before BBN and to achieve a successful gauge coupling unification. Is such hierarchy natural from a theoretical viewpoint? In this section, we address this question and show that a spectrum with $\ldc \ll \mH$ necessarily entails a fine tuning of parameters.~\footnote{A similar observation was made in Ref.~\cite{Harigaya:2016vda}.} For concreteness, we will perform our analysis on the $Q\+ \tilde D$ model with $SU(3)_{DC}$ dark color group and $5\+ 10$ GUT parent theory, although our conclusions will hold true for any model.

In the $Q\+ \tilde D$ model with $5\+ 10$ GUT theory one needs a hierarchy
\begin{equation*}
  m_{Q}, m_{\tilde{D}} \ll m_U, m_E, m_{\tilde{L}}
\end{equation*}
between the mass of $Q$, $\tilde D$ and that of their GUT partners $U$, $E,$ and $\tilde{L}$. We will assume $m_{Q}, m_{\tilde{D}}\sim m\lesssim \ldc$, and for simplicity, we take $m_U, m_E, m_{\tilde{L}} \sim \mH$. With a minimal content of GUT scalars given by one 24 plus one 5 of $SU(5)$, the GUT Lagrangian contains the following mass terms and Yukawa couplings:
\begin{equation}
  \label{eq:GUT510Lag}
  \begin{split}
    {\cal L} \supset  & - m_5 \bar\Psi_5 \Psi_5  -  m_{10}\bar\Psi_{10} \Psi_{10} - y_5 \bar\Psi_5 \phi_{24} \Psi_5 - y_{10} \bar\Psi_{10} \phi_{24} \Psi_{10} \\
    & - y_L \bar\Psi_5 P_L  \phi_{5}^{\dagger} \Psi_{10} -  y_R \bar\Psi_5 P_R \phi_{5}^{\dagger} \Psi_{10} + \text{h.c.}
\end{split}
\end{equation}
The tree-level expression of the dark quark masses is thus obtained by setting the scalar 24 to its vev, $\langle \phi_{24} \rangle = (\vgut/\sqrt{30})\, \diag(2,2,2,-3,-3)$.
We find
\begin{equation}
  \label{eq:masstree}
\begin{split}
  m_{\tilde{D}}& = m_5+\frac{2}{\sqrt{30}}y_5 \vgut\\
    m_{\T{L}}& =m_5-\frac{3}{\sqrt{30}}y_5 \vgut 
  \end{split}
  \hspace{1cm}
\begin{split}
 m_{U} & =m_{10}+\frac{2}{\sqrt{30}}y_{10}\vgut\\
    m_{E} & =m_{10}-\frac{3}{\sqrt{30}}y_{10}\vgut\\
    m_{Q} & =m_{10}-\frac{1}{2\sqrt{30}}y_{10 }\vgut\, .
    \end{split}
  \end{equation}
The hierarchy $m_{Q}, m_{\T{D}} \ll m_U, m_E, m_{\T{L}}$ can be obtained by taking both $m_{5, 10}$ and $y_{5, 10} \vgut$ of order $\mH$, and by performing two tunings: the first between $m_5$ and $y_5 \vgut$ to have $m_{\tilde D} \ll m_{\tilde L}$, the second between $m_{10}$ and $y_{10} \vgut$ to have $m_Q\ll m_{U}, m_E$. We see that there is no way to naturally split the masses of the light quarks from those of their GUT partners.

One could hope that theories with a non-minimal content of GUT scalars, having new Yukawa interactions, could lead to the desired hierarchy of masses.
An alternative strategy might be to take both $m_{5,10}$ and $y_{5,10} \vgut$ of order $m$ and try to generate the much larger contribution to $m_{U},m_E,m_{\tilde L}$ from higher-dimension operators.
To test these ideas, we have analyzed theories with additional GUT scalars transforming in the 75 and 200 of $SU(5)$ and computed the dark quark mass matrix by including the effect of $D=5$ operators. We focused on scalar fields in the 75 and 200 representations because these are the only ones, besides $\phi_{24}$, whose vev can break $SU(5)$ and preserve the SM gauge group, and whose presence in the Lagrangian is compatible with the perturbativity of the $SU(5)$ gauge coupling up to the Planck scale.
A simple analysis, reported in Appendix~\ref{appendix:splitting}, shows that splitting $Q$, $\tilde D$ from $U$, $E$, $\tilde L$ requires a particular alignment of the dark quark mass matrix in $SU(5)$ space. We find that none of the new Yukawa couplings and $D=5$ operators lead to the correct alignment. The details of our analysis are reported in Appendix~\ref{appendix:splitting}.  We conclude that generating the desired hierarchy necessarily entails a tuning of parameters.

The next question is whether such tuning is stable under radiative corrections.
Let us stick to a minimal GUT scalar content and consider for example the hierarchy $m_{\tilde D} \ll m_{\tilde L}$, which is obtained by tuning the values of $m_5$ and $y_5 \vgut$.
At energies above the GUT scale, mass terms $m_{5,10}$ and Yukawa couplings $y_{5,10}$ are multiplicatively renormalized.
This can be seen, for example, by using the spurionic transformation rules of mass terms and Yukawa couplings (see Tab.~\ref{table:spurion_charges}) under the $U(1)$ axial symmetries acting on $\Psi_5$ and $\Psi_{10}$:
\begin{table}[t]
\centering
\begin{tabular}{ | r | c | c | c | c | c | c | } 
\hline
  & $m_5$ & $m_{10}$ & $y_5$ & $y_{10}$ & $y_L$ & $y_R$  \\[0.1cm]
  \hline
 $U(1)_{5A}$   &  $-2$ & $0$ & $-2$ & $0$ & $+1$ & $-1$ \\ 
 $U(1)_{10A}$ &  $0$ & $-2$ & $0$ & $-2$ & $+1$ & $-1$\\
 \hline
\end{tabular}
\caption{Spurion charges assigned to mass terms and Yukawa couplings of the $5\+ 10$ GUT theory relative to the axial $U(1)_{5A}$ and $U(1)_{10A}$ symmetries defined by Eq.~(\ref{eq:U1axial}).}
\label{table:spurion_charges}
\end{table}
%
\begin{equation}
\label{eq:U1axial}
U(1)_{5A} : \ \Psi_5 \to e^{i \alpha\gamma_5} \Psi_{5}\, , \qquad
U(1)_{10A} : \ \Psi_{10} \to e^{i \beta\gamma_5} \Psi_{10}\, .
\end{equation}
%
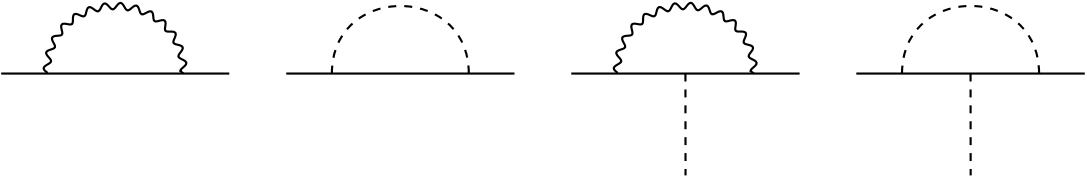
\begin{figure}[t]
  \centering
\begin{tikzpicture}
\begin{feynman} [scale=1.5, transform shape]
  \vertex(a1);
  \vertex [right= 0.4cm of a1] (a2);
  \vertex [right= 0.6cm of a2] (a3);
  \vertex [right= 0.6cm of a3] (a4);
  \vertex [right= 0.4cm of a4] (a5);
  \vertex [right = 2.5cm of a1] (aa1);
  \vertex [right= 0.4cm of aa1] (aa2);
  \vertex [right= 0.6cm of aa2] (aa3);
  \vertex [right= 0.6cm of aa3] (aa4);
  \vertex [right= 0.4cm of aa4] (aa5);
  \vertex [right = 5cm of a1] (aaa1);
  \vertex [right= 0.4cm of aaa1] (aaa2);
  \vertex [right= 0.6cm of aaa2] (aaa3);
  \vertex [right= 0.6cm of aaa3] (aaa4);
  \vertex [right= 0.4cm of aaa4] (aaa5);
  \vertex [below= 0.9cm of aaa3] (b3);
  \vertex [right = 7.5cm of a1] (aaaa1);
  \vertex [right= 0.4cm of aaaa1] (aaaa2);
  \vertex [right= 0.6cm of aaaa2] (aaaa3);
  \vertex [right= 0.6cm of aaaa3] (aaaa4);
  \vertex [right= 0.4cm of aaaa4] (aaaa5);
  \vertex [below= 0.9cm of aaaa3] (bb3);
  \diagram* [large] {
    (a1) --[thick]  (a5),
    (a2) -- [half left, looseness=1.7, thick, photon] (a4),
  };
  \diagram* {
    (aa1) --[thick]  (aa5),
    (aa2) -- [half left, looseness=1.7, thick, scalar] (aa4),
  };
\diagram* {
    (aaa1) --[thick]  (aaa5),
    (aaa2) -- [half left, looseness=1.7, thick, photon] (aaa4),
    (aaa3) -- [thick, scalar] (b3),
  };
  \diagram* {
    (aaaa1) --[thick]  (aaaa5),
    (aaaa2) -- [half left, looseness=1.7, thick, scalar] (aaaa4),
    (aaaa3) -- [thick, scalar] (bb3),
  };
\end{feynman}
\end{tikzpicture}
\caption{One-loop Feynman diagrams renormalizing $m_{5,10}$ and $y_{5,10}$ in the $Q\+ \tilde D$ model with $5\+ 10$ GUT theory, see Eq.~(\ref{eq:GUT510Lag}). Solid, dashed, and wavy lines denote respectively dark quarks, GUT scalars, and $SU(5)$ or $SU(3)_{DC}$ gauge bosons.}
\label{fig:1loop}
\end{figure} 
The one-loop corrections to $m_5$ and $y_5$ arise from the diagrams of Fig.~\ref{fig:1loop} and are schematically of the form
\begin{equation}
  \label{eq:dm}
\begin{split}
\delta m_5 & \propto \frac{g_5^2}{16\pi^2} m_5 + \frac{g_{DC}^2}{16\pi^2} m_5 + \frac{|y_5|^2}{16\pi^2} m_5 + \frac{y_L^* y_R}{16\pi^2} m^*_{10} \\[0.2cm]
\delta y_5 & \propto \frac{g_5^2}{16\pi^2} y_5 + \frac{g_{DC}^2}{16\pi^2} y_5+ \frac{|y_5|^2}{16\pi^2} y_5 + \frac{y_L^* y_R}{16\pi^2} y^*_{10}\, ,
\end{split}  
\end{equation}  
where only the dependence on mass scales and coupling constants has been shown, neglecting $O(1)$ coefficients. If too large, such corrections spoil the tree-level tuning of parameters, thus destabilizing the mass hierarchy. The largest contribution in Eq.~(\ref{eq:dm}) comes from the first term, due to loops of $SU(5)$ gauge bosons.~\footnote{It is not difficult to make the contributions from Yukawa couplings (last two terms in (\ref{eq:dm})) small enough. For example, since $y_{5,10} \sim \mH/\vgut$, the third term in the first line of~Eq.~(\ref{eq:dm}) is smaller than $m$ for $\mH \lesssim (16\pi^2 m \vgut^2)^{1/3}\sim  1\times 10^{13}\,$GeV, where for the last estimate we assumed $m\sim \ldc \sim 100\,$TeV and $\vgut \sim 10^{16}\,$GeV. Furthermore, one can assume $y_L y_R \lesssim 16\pi^2 m/\mH$ so that also the fourth term in the first line of~Eq.~(\ref{eq:dm}) is smaller than $m$.}
This is smaller than $m$ for $m\ll \mH \lesssim (4\pi/\alpha_{GUT}) m$; for example, assuming $m \sim \ldc \sim 100\,$TeV and $\alpha_{GUT} \sim 1/15$ (see Fig.~\ref{fig:delta_cartoon}) gives $10^5\,\text{GeV} \ll \mH \lesssim 2 \times 10^7\,$GeV. This is a rather narrow range for $\mH$, which lies well below the estimated lower bound from BBN constraints (see Sec.~\ref{subsec:degGUT}). These considerations therefore show that the required mass hierarchy between light dark quarks and their GUT partners is not stable under radiative corrections.

It is interesting to notice that loop corrections below the GUT scale do not destabilize the hierarchy $m\ll \mH$ in the $Q\+ \tilde D$ model. This is because there exist no Yukawa interactions that mix light ($Q$, $\tilde D$) with heavy ($U$, $E$, $\tilde L$) dark quarks in the effective theory below $\mgut$. Light dark quark masses are thus multiplicatively renormalized, as it can be verified through a spurion argument in terms of the axial $U(1)$ symmetry acting (only) on light dark quark fields. Therefore, radiative corrections that destabilize the hierarchy only arise from the domain of energies above the GUT scale. This is, however, a peculiar feature of the $Q\+ \tilde D$ model, which does not hold true in other models, such as $L\+ E$.

Besides radiative corrections, also contributions from higher-dimensional operators generated at the Planck scale can destabilize the hierarchy $m\ll \mH$. For example, $D=5$ operators of the form
\begin{equation}
\frac{c}{M_P} \bar{\Psi}_5 \phi_{24}^2 \Psi_5\, , \qquad \frac{c}{M_P} \bar{\Psi}_{10} \phi_{24}^2 \Psi_{10}
\end{equation}
imply corrections $\delta m_{5,10} \sim c \,\vgut^2/M_P$. For $c\sim 1$ (and assuming $\vgut \sim 10^{16}\,$GeV) these are of order $10^{14}\,$GeV, hence much larger than $\ldc \sim 100\,$TeV. Turning the argument around, one concludes that $D=5$ operators do not destabilize the mass hierarchy if their coefficients are sufficiently small, $c \lesssim 10^{-9}$.
Incidentally, this same estimate shows that values of $\mH$ much smaller than $10^{14}\,$GeV can be considered as technically natural~\footnote{One can make again a spurion argument in terms of the axial symmetries of Eq.~(\ref{eq:U1axial}).} only as long as the GUT theory has a weakly-coupled UV completion at the Planck scale so that higher-dimensional operators have sufficiently small coefficients.

To conclude, the arguments presented above show that theories where light dark quarks are much lighter than their GUT partners
are fine tuned. In the $Q\+ \tilde D$ model, obtaining $m_U, m_E, m_{\tilde L}$ all of order $M$ and much larger than $m_Q$, $m_{\tilde D}$ requires two tunings. Clearly, realizing a split, rather than an almost degenerate, spectrum of GUT partners implies additional tunings.

\section{Summary and Outlook}
\label{sec:summary_cosmo}

In this work, we have systematically studied and classified $SU(5)$-GUT completions of accidentally composite dark matter models.
These theories postulate new dark quarks transforming in vectorlike or real representations of an $\sundc$ or $\sondc$ dark color group and of the $SU(5)$ grand unification group. Dark Matter is predicted in the form of an accidentally stable dark baryon.
A first systematic classification of accidental composite DM models was performed in Ref.~\cite{Antipin:2015xia}, where it was shown that dark fermion $SU(5)$ irreps must split into light dark quarks,  whose bound states include the DM candidate, and their much heavier GUT partners. Such a mass hierarchy is required mainly to avoid unwanted cosmologically stable states in the theory, i.e. states whose quantum numbers under the SM gauge group are not compatible with present experimental bounds. Furthermore, a split spectrum can improve the quality of gauge coupling unification. Following Ref.~\cite{Antipin:2015xia},  we assumed that light dark quarks have masses at or below the dark color dynamical scale $\ldc$. In such a framework, the thermal relic abundance of the lightest accidentally stable dark baryon can reproduce the DM density for values of $\ldc \sim 100\,$TeV. The DM particle itself has a mass of order $100\,$TeV, and must have vanishing hypercharge to comply with direct detection bounds.

The authors of Ref.~\cite{Antipin:2015xia} performed a classification of viable models focussing on the DM sector of light dark quarks. In this work, we have adopted an $SU(5)$-GUT perspective and selected only theories where all unwanted species symmetries of the DM sector can be broken by higher-dimensional operators generated at the GUT scale or at the scale of the dark quark GUT partners. We find that higher-dimensional operators generated at the Planck scale are too suppressed to let metastable dark pions decay before the onset of BBN. This means that a viable GUT theory must possess no accidental species numbers. Our selection is outlined in Sect.~\ref{sec:classification} and the results are reported in Appendix~\ref{appendix:models_list}. We reproduce all the golden class models of Ref.~\cite{Antipin:2015xia}, but find a different list of silver class models.

As a second step, we analyzed the unification of SM gauge couplings for each of the selected theories.
Assuming that all the dark quark GUT partners have a similar mass, we find that very few theories can improve or just equal the quality of gauge coupling unification in the SM. Specifically, only the $Q\+\tilde D$ model with $10\+5$ or $15\+5$ GUT parent theories leads to a more precise unification. This is mainly because, as already noticed in previous literature, one-step unification requires new colored particles, and only few selected DM models have colored dark quarks. One of the reasons for such a scarcity of models is that $D\geq 6$ operators generated at the dark quark GUT partner scale preserve color species symmetries (see Sec.~\ref{subsec:longlived}). As a consequence, a viable model cannot have both dark quarks with only electroweak quantum numbers and colored dark quarks. It is interesting to notice that adding new dark scalar fields enlarges the spectrum of viable theories, since unwanted species symmetries can be broken by means of new Yukawa couplings (see Ref.~\cite{Palmisano:2024mxj}). Good unification can be achieved also in models with color-singlet dark quarks if one allows for a non-degenerate spectrum of GUT partners (see Fig.~\ref{fig:nondeg}).

The hierarchy in the mass spectrum of dark quarks emerges as a key ingredient of our class of theories, and it is therefore legitimate to ask if it can be obtained naturally. A simple analysis shows (see Sec.~\ref{sec:mass_split}) that splitting light dark quarks from their GUT partners requires a tuning of parameters that is generally unstable under radiative corrections and the effect of Planck-suppressed operators. This observation clearly disfavors models with a split spectrum of GUT partners, since these require even more fine tunings.
One should keep in mind, however, that a hierarchical spectrum might result from specific UV completions. 
Extra dimensional orbifold GUTs, for example, could be a possible framework to address this problem.

If one assumes that the UV theory can explain only the hierarchy between light dark quarks and their GUT partners, while the latter have an almost degenerate spectrum, then our analysis selects only two models: $Q\+\tilde D$ with $SU(3)_{DC}$ dark color group and $10\+5$ or $15\+5$ GUT parent theory; and $N$ with $SO(3)_{DC}$ dark color group and $24$ GUT parent theory. Interestingly, the $Q\+\tilde D$ model with $10\+ 5$ parent theory allows for a perfect gauge coupling unification. This justifies a thorough analysis of the phenomenology and cosmology of such a theory, which will be presented in a forthcoming paper~\cite{BCV}.

In this work we have focused on the case of light dark quark masses below the dark color scale, $m\lesssim \ldc$. A natural extension of our study would be to consider theories with $m > \ldc$. A thorough analysis in this direction was performed by Ref.~\cite{Mitridate:2017oky}, which focused on the cosmological history dictated by the light dark quark degrees of freedom.
It would be interesting to study, in particular, the impact that GUT partners have on phenomenology and cosmology.
Values of $m$ much larger than $100\,$TeV pose the problem of the accidental stability of the DM: the higher the DM mass, the higher the dimension of operators breaking dark baryon number should be. Furthermore, the results of Refs.~\cite{Mitridate:2017oky,Contino:2018crt} suggest that such large mass values might be excluded by a too slow glueball decay if the DM has a thermal abundance.
Alternatively, viable theories could be found on a branch with $m \sim 10^4-10^6\,$GeV and $\ldc \sim 10^2 - 10^5\,$GeV characterized by the absence of an epoch of early matter dominance. This picture, sketched in Ref.~\cite{Mitridate:2017oky} and supported by the results of Ref.~\cite{Contino:2018crt}, does not, however, take into account the role of the dark quark GUT partners.  These can have an impact on both the selection of viable theories and their cosmology. An even more drastic departure from the scenario of Ref.~\cite{Mitridate:2017oky} would be implied by theories in which light dark quarks have very large masses,
$m \gtrsim 10^9\,$GeV, and the DM abundance is set before the end of reheating~\cite{Harigaya:2016vda}. Dark Matter candidates with non-vanishing hypercharge are acceptable in that case, opening the possibility for a new class of superheavy accidental DM theories. We leave the study of these interesting directions to a future work.

\section*{Acknowledgments}
We would like to thank Graham Kribs, Andrea Luzio, Andrea Mitridate, Alessandro Podo, Michele Redi, and Ennio Salvioni for their comments, discussions, and clarifications on their work. SV also wishes to thank the DESY Theory Group and the High-Energy Physics Group at Sapienza for their hospitality during a significant portion of this work. SB is supported by the Israel Academy of Sciences and Humanities and the Council for Higher Education’s Excellence Fellowship Program for International Postdoctoral Researchers.

\appendix
\clearpage
\section{Classification Results: Candidate Models}
\label{appendix:models_list}

In this Appendix, we list all the candidate models that pass the requirements discussed in Section~\ref{sec:classification} and show their GUT parent theories.

\subsection{SU$(N)_{DC}$ Models}
We start by considering $\sundc$ models. Each row in the tables below gives a viable model passing our model-building criteria, the allowed number of dark colors $\ndc$, the DM candidates, and the minimal parent GUT theories.

\subsubsection*{Models with $T$}

The perturbativity condition on the hypercharge coupling sets an upper bound on the number of dark colors, $\ndc<5$. Furthermore, for $\ndc=3$, $SU(2)_L$ and $U(1)_Y$ perturbativity forbids the presence of more than one electroweak triplet, hence combinations with $V$, $\tilde{T}$, and $E\+\tilde{E}$ do not work as low-energy models. 
Moreoever, the following $SU(3)_{\text{DC}}$ models
\begin{equation}
    T,\quad T\+\tilde{L},\quad T\+E,\quad T\+\tilde{E},\quad T\+\tilde{L}\+E
\end{equation}
contain no DM candidate and are not further considered. Finally, for $\ndc=4$ only $N$ can be added.

\begin{center}
\begin{tabular}{  c | c | c | c}
\textbf{Model} & Range of $\ndc$ & DM candidates &  Parent theories \\
\hline &&& \\[-0.4cm]
$T\oplus N$  &    $\ndc=3,4$  & $N^{\ndc}$ & $15 \+ 5 \, \+ (1 \,\text{or}\, 24)$\\[0,1cm]
$T\oplus L \,(\oplus N)$  &    $\ndc=3$  & $TLL\, (NNN) $ & $\overline{5} \+ 15\+ 5  \+ (1 \,\text{or}\, 24)$\\[0,1cm]
$T\oplus \tilde{L} \oplus N$  &    $\ndc=3$  & $ NNN $ &$5\+15\+(1 \,\text{or}\, 24)$\\[0,1cm]
$T\oplus E \oplus N$  &    $\ndc=3$  & $ NNN $ &$10\+15\+ 5 \, \+ (1 \,\text{or}\, 24)$\\[0,1cm]
$T\oplus \tilde{E} \oplus N$  &    $\ndc=3$  & $ NNN $ &$\overline{10}\+15\+\overline{5}\+  5 \, \+ (1 \,\text{or}\, 24)$\\[0,1cm]
$T\oplus L\+\tilde{L} \, (\oplus N)$ &    $\ndc=3$  & $TLL\, (NNN,\, NL\tilde{L}) $  & $\overline{5} \+ 15 \+  5 \, \+ ( 1 \,\text{or}\, 24)$ \\[0,1cm]
$T\oplus L\+ E\, (\oplus N)$  &    $\ndc=3$  & $TLL,\, LLE \, (NNN) $ & $\overline{5} \+ 10\+ 15 \+ 1\+  5$\\[0,1cm]
$T\oplus L\+ \tilde{E}\, (\oplus N)$  &    $\ndc=3$  & $TLL \, (NNN) $ & $\overline{5} \+ \overline{10}\+ 15 \+ 1\+  5$\\[0,1cm]
$T\oplus \tilde{L}\+ E\oplus N$  &    $\ndc=3$  & $NNN $ & $15\+ 5\+ 10\+ (1 \,\text{or}\, 24)$\\[0,1cm]
$T\oplus \tilde{L}\+ \tilde{E}\, (\oplus N)$  &    $\ndc=3$  & $\tilde{L}\tilde{L}\tilde{E}\, (NNN)$ & $15\+ 5\+\overline{10} \+ 1\+  \overline{5}$\\[0,1cm]
\end{tabular}
\end{center}
For the model $T \oplus L$, the operator $\bar{L}THHH$ can be generated by integrating out GUT partners $\tilde{L}$ and $N$, hence the GUT embedding $15 \oplus \bar{5} \oplus 5 \oplus 24$.

\subsection*{Models with $V$}

$V^{N_{DC}}$ is always a good DM candidate for any $\ndc$. In the third column of the table below, we list any additional DM candidates in the theory. Dark Matter candidates are written as gauge singlet components where the indices $n, \; m, \; l, \; h$ can take any integer values from 0 and should sum up to the total number of dark colors $\ndc$.

\begin{center}
\begin{tabular}{  c | c| c |c}
\textbf{Model} & Range of $\ndc$ & DM candidates &  Parent theories \\
\hline &&& \\[-0.4cm]
$V  \,(\oplus N) $ &  $\ndc\leq 4$   & $V^n N^m$ &$24 \+ \bar{5}$ \\ [0.1cm]
$V\+ L \,(\oplus N)$      &    $\ndc=3$  &  $V^n N^m$  & $24 \+ \overline{5} $\\ [0.1cm]
$V\+ E \,(\oplus N)$      &    $\ndc\leq 4$       &  $V^n N^m$  & $24 \+  10 \, \+ \, 5$\\ [0.1cm]
$V\+ L\+ \tilde{L} \,(\oplus N)$  &    $\ndc=3$  &  $V^n N^m (\tilde{L}L)^l$ &  $24 \+ \overline{5} \+ 5$\\ [0.1cm]
$V\+ L\+ E \,(\oplus N)$  &    $\ndc=3$  &  $V^n N^m (LLE)^l$ &  $24 \+ \overline{5} \+ 10 \, \+ \, 5$\\ [0.1cm]
$V\+ L\+ \tilde{E} \,(\oplus N)$  &    $\ndc=3$  &  $V^n N^m$ &  $24 \+ \overline{5} \+ \ov{10} $\\ [0.1cm]
$V\+ E\+ \tilde{E} \,(\oplus N)$  &    $\ndc\leq 4$  &  $V^n N^m(\tilde{E}E)^l$ &  $24 \+ \ov{10} \+ 10 \, \+ \, 5 \+ \bar{5}$\\ [0.1cm]
$V\+ L\+\tilde{L}\+ E \,(\oplus N)$  &   $\ndc=3$  &  $V^n N^m(\tilde{L}L)^l$ &  $24 \+ \overline{5} \+ 5 \+  10$\\ [0.1cm]
$V\+ L\+E\+ \tilde{E} \,(\oplus N)$  &    $\ndc=3$ &  $V^n N^m(\tilde{E}E)^l$ &  $24 \+ \overline{5}  \+ \ov{10} \+ 10 \, \+ \, 5  \+ (1 \, \mathrm{or} \, 24)$\\ [0.1cm]
$V\+L\+\tilde{L}\+ E\+ \tilde{E} \,(\oplus N)$  &   $\ndc=3$  &  $V^n N^m(\tilde{L}L)^n(\tilde{E}E)^h$ &  $24 \+ \overline{5} \+ 5 \+ \ov{10} \+ 10$\\ [0.1cm]
\end{tabular}
\end{center}

Note that the models $V$ and $V \+ N$ with 24 parent GUT theory are not viable. Indeed, since the 24 is real, the GUT theory with only this representation is invariant under G-parity. 
In this case, G-parity violating operators cannot be generated at or below $\mgut$, and mesons $\bar{V}V$ and $\bar{V}N$ are too long-lived. Adding the complex $SU(5)$ representation $\bar{5} $ to the GUT field content breaks G-parity invariance and makes the theory viable.

\subsubsection*{Models with $E$}

Models $E$ and $E\+\tilde{L}$ have no DM candidate and therefore are not viable.
In the table below, DM candidates are written as gauge singlet components where the indices $n, \; m, \; l, \; h$ can take any integer values from 0 and should sum up to the total number of dark colors $\ndc$.\\

\begin{center}
\begin{tabular}{  c | c| c |c}
\textbf{Model} & Range of $\ndc$ & DM candidates &  Parent theories \\
\hline &&& \\[-0.4cm]
$E\+ N$  &    $\ndc\leq 13$  & $N^{\ndc}$  & $10\+(1 \,\text{or}\, 24) \+ \, 5$ \\[0.1cm]
$E\+ L \, (\+ N)$  &    $\ndc\leq 9$  & $(LLE)^nN^m $  & $\ov{5} \+ 10 \+ 5 \+ (1 \,\text{or}\, 24)$  \\[0.1cm]
$E\+ \tilde{L}\+ N$  &    $\ndc\leq 9$  & $N^{\ndc}$  & $10\+5\+(1 \,\text{or}\, 24)$ \\[0.1cm]
$E\+ \tilde{E} \,(\+ N)$  &    $\ndc\leq 6$  & $(E\tilde{E})^nN^m$  & $10\+\ov{10} \+ 5 \+ \bar{5} \, \+(1 \,\text{or}\, 24)$ \\[0.1cm]
$E\+ L\+\tilde{L} \,(\+ N)$  & $\ndc\leq 6$ &$(E\tilde{E})^n(LLE)^mN^l$ & $10 \+ 5 \+ \ov{5} \, \+(1 \,\text{or}\, 24)$\\ [0.1cm]
$E\+ L\+\tilde{E} \,(\+ N)$  & $\ndc\leq 5$ &$(L\tilde{L})^n(LLE)^mN^l$ & $10 \+ \ov{5} \+ \ov{10} \, \+ \, 5 \, \+ \,(1 \,\text{or}\, 24)$\\ [0.1cm]
$E\+ L\+\tilde{E}\+\tilde{L} \,(\+ N)$  & $\ndc\leq 4$ &$(E\tilde{E})^n(L\tilde{L})^m(LLE)^lN^h$ & $10 \+ \ov{5} \+ \ov{10}\+ 5 \, \+ \, 5 \, \+ \,(1 \,\text{or}\, 24)$\\ [0.1cm]
\end{tabular}
\end{center}
For the model $E\+ \tilde{E}$, the generation of $\bar{E}\tilde{E} (H^{c \dagger}D_\mu H)^2$ operator requires integrating out GUT partners $\T{L}, L$ and $N$, resulting in the minimal GUT parent theory $10 + \bar{10} + 5 + \bar{5} + 24$.

\subsubsection*{Models with $L$}

The model $L$ has no DM candidate. In the table below, the indices $n$, and $m$ are integers whose sum equals the number of dark colors $\ndc$.\\

\begin{center}
\begin{tabular}{  c | c| c |c}
\textbf{Model} & Range of $\ndc$ & DM candidates &  Parent theories \\
\hline &&& \\[-0.4cm]
$L\+ N$ & $\ndc\leq 18$ & $N^{\ndc}$ & $\ov{5}\+ (1 \,\text{or}\, 24)$\\ [0.1cm]
$L\+ \tilde{L} \,(\+ N)$ & $ \ndc\leq 9$ & $(L\tilde{L})^nN^m$ &  $\ov{5}\+ 5\+ (1 \,\text{or}\, 24)$\\ [0.1cm]
\end{tabular}
\end{center}

\subsubsection*{Coloured dark fermion models }

Viable DM models can be built only out of combinations of $Q,\,\tilde{U},\,\tilde{D}$. No DM candidate exists for $\tilde{U}\+ Q$, since the hypercharges of these two dark quarks have the same sign. We are thus left with the following possibilities:
\begin{center}
\begin{tabular}{  c | c| c |c}
\textbf{Model} & Range of $\ndc$ & DM candidates &  Parent theories \\
\hline &&& \\[-0.4cm]
$Q\+ \tilde{D}$ & $\ndc = 3$ & $QQ\tilde{D}$ & $5\+ 10$, $5\+ 15$\\ [0.1cm]
$Q\+ \tilde{D}\+ \tilde{U}$ & $ \ndc = 3$ & $QQ\tilde{D},\, \tilde{D}\tilde{D}\tilde{U}$ &  $5\+ 10\+ \ov{10}$\\ [0.1cm]
$\tilde{D}\+ \tilde{U}$ & $ \ndc =3, \; 6$ & $\tilde{D}\tilde{D}\tilde{U}, \; \tilde{D}\tilde{D}\tilde{U}\tilde{D}\tilde{D}\tilde{U} $ &  $5\+ 10\+ \ov{10}$\\ [0.1cm]
\end{tabular}
\end{center}
Notice that for the $\tilde{D} \, \+ \, \tilde{U}$ model, dark colour confinement and perturbativity allow $\ndc \leq 8$. The further requirement of an SM singlet DM candidate selects $\ndc= 3,6$. Similarly, only $\ndc =3$ ensures to have dark color confinement, perturbativity, and an SM singlet DM candidate in the models $Q \+ \tilde{D}$ and $Q \+ \tilde{D} \+ \T{U}$.

\subsection{SO$(N)_{DC}$ Models}

We now move to $\sondc$ models.
In this case, baryons can be constructed by taking a completely antisymmetric dark color contraction of both $\q$ and $\q^c$ kinds of fields, since the fundamental of $\sondc$ is real. We recall that, for $\sondc$ theories, we do not impose the constraint $\ndf<N_\text{conf}$ from dark color confinement.

\subsubsection*{Models with $T$}

Dark color asymptotic freedom requires $N_{DC}>3$, however the request of SM hypercharge perturbativity forces $N_{DC}<5$. Hence, only $\ndc=4$ is allowed. 
\begin{center}
  \begin{tabular}{  c | c | c| c}
\textbf{Model} & Range of $\ndc$ & DM candidates &  Parent theories \\
    \hline &&& \\[-0.4cm]
$T$ &     $\ndc=4$  & $(TT^c)^2$ & $15 \+ 5 \+(1 \,\text{or}\, 24)$ \\[0.1cm]
$T\oplus N$  &    $\ndc=4$  & $(TT^c)^2,\, TT^cNN $ & $15 \+  \, 5 \, \+ (1 \,\text{or}\, 24)$
\end{tabular}
\end{center}

\subsubsection*{Models with $V$}

$V^{N_{DC}}$ is always a good DM candidate. The third column of the table below lists any additional DM candidates. The indices $n$, $m$, $l$ and $h$ are integers whose values start from~0 and sum to $\ndc$.
\begin{center}
\begin{tabular}{  c | c| c |c}
\textbf{Model} & Range of $\ndc$ & DM candidates &  Parent theories \\
\hline &&& \\[-0.4cm]
$V$ &  $3 \leq \ndc\leq 9$   & $V^{\ndc}$ &  $24$ \\ [0.1cm]
$V\+ L$  &    $4\leq\ndc\leq 6$  & $(LL^c)^n V^m$ & $ 24 \+ \bar{5}$\\[0.1cm]
$V\+ E$         & $ 3 \leq \ndc\leq 9$                & $(EE^c)^n V^m$ &  $24 \+ 10 \, \+ \, \bar{5}$\\ [0.1cm]
$V\+ E\+ L$       & $4\leq\ndc\leq 6$         & $(LL^c)^n(EE^c)^m V^l$ & $ 24 \+ 10 \+ \bar{5}$\\ [0.1cm]
$V\+ N$ & $ 3 \leq \ndc\leq 9$ & $V^n N^m$ &  $24$ \\ [0.1cm]
$V\+ L\+ N$  & $4\leq\ndc\leq 6$ & $(LL^c)^n V^m N^l$ &  $24\+ \bar{5}$\\ [0.1cm]
$V\+ E\+ N$  & $ 3  \leq\ndc\leq 9$ & $(EE^c)^n V^m N^l$ & $ 24\+ 10 \+ \bar{5} $ \\ 
$V\+ E\+ L\+ N$  &  $4\leq\ndc\leq 6$ & $(LL^c)^n(EE^c)^m V^lN^h$  & $24 \+ 10 \+ \bar{5}$\\ 
\end{tabular}
\end{center}

\subsubsection*{Models with $E$}

The viable models with $E$ are reported in the table below (the indices $n$, $m$, and $l$ are integers whose values start from~0 and sum to $\ndc$). In the $E$ model, a viable DM candidate exists only for $\ndc$ even.
\begin{center}
\begin{tabular}{  c | c | c| c }
\textbf{Model} & Range of $\ndc$ & DM candidates &  Parent theories \\
  \hline &&& \\[-0.4cm]
$E$ & $4 \leq\ndc\leq 12$ & $(EE^c)^{\ndc/2}$ & $10 \+ \bar{5}\+ (24\, \text{or}\, 1)$ \\ [0.1cm]
$E\+ L$  &    $4\leq\ndc\leq 9$  & $(EE^c)^n (LL^c)^m(LLE)^l$  &  $10\+ \bar{5} \+ (24\, \text{or}\, 1)$ \\[0.1cm]
$E\+ N$ & $3\leq \ndc\leq 13$ & $(EE^c)^n N^m$ & $10 \+ \bar{5}\+ (24\, \text{or}\, 1)$ \\ [0.1cm]
$E\+ L\+ N$  & $4\leq\ndc\leq 9$ &$(EE^c)^n (LL^c)^m N^l (LLE)^k$ & $10 \+ \bar{5}\+ (24\, \text{or}\, 1)$  
\end{tabular}
\end{center}

\subsubsection*{Models with $L$}

In the table below, the indices $n$ and $m$ are integers whose values start from~0 and sum to $\ndc$.
In the $L$ model, a viable DM candidate exists only for $\ndc$ even. 
\begin{center}
\begin{tabular}{  c | c | c |c}
\textbf{Model} & Range of $\ndc$ & DM candidates &  Parent theories \\
\hline &&& \\[-0.4cm]
$L$  &    $4\leq\ndc\leq 18$  & $(LL^c)^{\ndc/2}$ & $\bar{5}  \+ (24\, \text{or}\, 1)$ \\[0.1cm]
$L\+ N$ & $3\leq \ndc\leq 18$ & $(LL^c)^nN^m$ &  $\bar{5} \+ (24\, \text{or}\, 1)$\\ [0.1cm]
\end{tabular}
\end{center}

\subsubsection*{Colored dark fermion models}

From the list of symmetry-breaking operators of Tab.~\ref{table:SU(N)operators}, one can see that in no model with $U,\, D,\, Q$ all the species symmetries can be broken.
For example, consider the $SO(3)_{DC}$ model $Q\+D$ with DM candidate $QQD^c$: in the list of Tab.~\ref{table:SU(N)operators}, the Yukawa operator $\bar QD^cH$ breaks one linear combination of the two $U(1)$ species symmetries acting on $Q$ and $D$, while no other operator breaks the orthogonal combination. The lightest dark meson among $\bar Q Q^c$, $\bar D D^c$, $\bar Q D$ is thus metastable and decays after $1\,$s, making the model not viable.
In models with $S$, dark color and $SU(3)_c$ perturbativity are never both satisfied. Hence, the only representation allowed is $G$, which leads to the following models:
\begin{center}
\begin{tabular}{  c | c | c|c}
\textbf{Model} & Range of $\ndc$ & DM candidates &  Parent theory \\
\hline &&& \\[-0.4cm]
$G \,(\+ N)$  &    $\ndc=4,5$  & $G^n N^m$ &  $24$\\[0.1cm]
\end{tabular}
\end{center}
Here again, $n$ and $m$ are integers whose values run from 0 and sum to $\ndc$.
\section{Mass Splitting}\label{appendix:splitting}

In this appendix, we elaborate more on the fine tuning required to split light dark quarks from their GUT partners. We analyze, in particular, the effect of higher-dimensional operators and consider a non-minimal content of GUT scalar fields. For concreteness, we work in the $Q\+ \tilde D$ model with $SU(3)_{DC}$  dark color group and $5\+ 10$ GUT parent theory, but the conclusions that we obtain are more general.

The expression of the dark quark masses at tree level and for a minimal content of GUT scalar fields was given in Eq.~(\ref{eq:masstree}) and shows clearly that splitting $Q$, $\tilde D$ from $U$, $E$ and $\tilde L$ requires fine tuning of parameters. It is simple and useful to generalize this result. To this aim, we consider the most general structure that dark quark masses can have in the presence of $SU(5) \to SU(3)\times SU(2)\times U(1)_Y$ spontaneous breaking. For example, the mass matrix of $\tilde D$, $\tilde L$ can be written in terms of two $SU(5)$ structures corresponding to the decomposition $\bar{5} \times 5 = 1 + 24$:
\begin{equation}
  \mathcal{M}_5 = m_1 \cdot \mathbb{I} + m_{24} \cdot T_{24}\, ,
\end{equation}
where $\mathbb{I}$ is the $5\times 5$ identity matrix, and the matrix $T_{24}$ is aligned with the SM-preserving vev of the scalar 24: $\langle \Phi_{24} \rangle = \vgut T_{24}$. It follows:
\begin{equation}
  \label{eq:massDL}
    \begin{split}
     m_{\tilde{D}} &=m_1 + \frac{2}{\sqrt{30}} m_{24} \\
      m_{\T{L}}&=m_1 - \frac{3}{\sqrt{30}}m_{24}\, .
    \end{split}
\end{equation}
The mass parameters $m_1$ and $m_{24}$ encode all possible contributions to the masses of $\tilde D$ and $\tilde L$, including those from higher-dimensional operators and radiative corrections. Some of the corresponding Feynman diagrams are shown in Fig.~\ref{fig:1loop}.
At tree level one has $m_1 = m_5$, $m_{24}= y_5\vgut$, compatibly with Eq.~(\ref{eq:masstree}).

Similarly, the mass matrix for dark quarks in the $10$ is written in terms of three structures ($\bar{10} \times 10 = 1 + 24 + 75$):
\begin{equation}
  \mathcal{M}_{10} = m_1 \cdot \mathbb{I} + m_{24} \cdot T_{24} + m_{75} \cdot T_{75}\, ,
\end{equation}
where $\mathbb{I}$, $T_{24}$, and $T_{75}$ are conveniently expressed as tensors with two covariant and two contravariant indices.
The tensor $T_{75}$ is aligned with the SM-singlet component of a 75 irrep of $SU(5)$ (see Eq.~(\ref{eq:75_vev})).
It follows:
\begin{equation}
  \label{eq:massQ}
\begin{split}
 m_{U} & =m_{1}+\frac{2}{\sqrt{30}} m_{24} + \frac{2}{\sqrt{72}}\, m_{75}\\
 m_{E} & =m_{1}-\frac{3}{\sqrt{30}} m_{24} +  \frac{6}{\sqrt{72}}\,m_{75}\\
 m_{Q} & =m_{1}-\frac{1}{2\sqrt{30}} m_{24}  - \frac{2}{\sqrt{72}}\, m_{75}\, .
\end{split}
\end{equation}
At tree level, in a theory with only $\phi_5$ and $\phi_{24}$, one has $m_1=m_{10}$, $m_{24} = y_{10}\vgut$, $m_{75} =0$, compatibly with Eq.~(\ref{eq:masstree}).
In general, the $SU(5)$ structures in the mass matrix ${\cal M}_R$ of a representation $R$ of $SU(5)$ are determined by the $SU(5)$ irreps in the tensor product $\bar R\times R$ which contain a SM singlet.

Equations~(\ref{eq:massDL}) and~(\ref{eq:massQ}) show that splitting $Q$, $\tilde D$ from $U$, $E$ and $\tilde L$ requires a particular alignment of the dark quark mass matrices in $SU(5)$ space.
For example, solving Eq.~\eqref{eq:massDL} in terms of $m_1$ and $m_{24}$, and assuming $m_{\T{D}} \ll m_{\T{L}}$ gives:
\begin{equation}
  \label{eq:fine_tuning}
     \frac{\sqrt{30} \ m_1}{m_{24}} = - 2  -5 \cdot \frac{m_{\T{D}}}{m_{\T{L}}} + O\!\left( \frac{m_{\T{D}}^2}{m_{\T{L}}^2}\right)\, .
\end{equation}
Radiative corrections contribute to both $m_1$ and $m_{24}$ and do not satisfy the above relation naturally. Furthermore, it turns out that none of the new Yukawa interactions allowed with a non-minimal content of GUT scalar fields, and none of the $D=5$ operators leads to the alignment of Eq.~(\ref{eq:fine_tuning}).  This is shown by the following analysis.


We consider new GUT scalar fields in higher-dimensional $SU(5)$ irreps
besides $\phi_5$ and $\phi_{24}$. We focus on scalar fields whose vev
breaks $SU(5)$ but is invariant under the SM gauge group. We require
that the addition of any such new GUT scalar does not drive the
$SU(5)$ coupling strong below the Planck scale. This puts a constraint
on the Dynkin index of the scalar's representation $R_s$:
\begin{equation} \label{eq:largest_rep}
  T(R_s) \leq 6\left(\frac{101}{6}-\frac{4}{3}N_{DC}\sum_f a_f T(R_f)+2\pi \frac{\alpha_{\text{GUT}}^{-1}}{\log\left(\frac{M_{\text{Pl}}}{\mgut}\right)}\right)\, .
\end{equation}
Here $T(R)$ denotes the Dynkin index of the representation $R$, the sum runs over all $SU(5)$ fermion representations, and $a_f = 1, 1/2$ respectively for Dirac and Majorana fermions. Note that $\alpha_{\text{GUT}}$ depends on the value of the heavy dark quark mass $\mH$. The first contribution comes from the running of SM fields, GUT gauge bosons, and the colored scalar in $\phi_5$, whereas the second term includes the running from dark quarks. For the $SU(3)_{DC}$ $Q \+ \T{D}$ model in $5 \+ 10$, the second term gives a value of 8.
The highest allowed value of $ T(R_s)$ for this model is 245 (where we have used $\mh = 10^{11} \ \mathrm{GeV}$ for perfect unification at $\mgut \approx 10^{17} \ \mathrm{GeV}$ with  $\alpha_{\text{GUT}}^{-1} \approx 16$). The only $SU(5)$ irreps that fulfill  Eq.~(\ref{eq:largest_rep}) and contain a SM singlet are: 200 ($T(200) = 100$), 75 ($T(75) = 25$), and 24 ($T(24)= 5$).

\begin{itemize}
\item $\mathbf{\phi_{24}}$:
\newline
The $D=5$ operators that one can write in terms of $\phi_{24}$ are (here and in the following, we use a simplified notation and denote $(\Psi_5)^i\to \Psi^i$, $(\Psi_{10})^{ij}\to\Psi^{ij}$, $(\phi_{24})^i_j\to \phi^i_j$):
    \begin{equation}\label{eq:su_5_L}
    \begin{split}
       & \frac{c_1}{\Lambda} \bar\Psi_{i} \Psi^i\phi^m_n\phi^n_m + \frac{c_2}{\Lambda}\bar\Psi_{ij}\Psi^{ij}\phi^m_{n}\phi^n_{m}+\frac{d_1}{\Lambda}\bar\Psi_{i}\Psi^j\phi^i_{n}\phi^n_{j} \\[0.2cm] 
      & +\frac{d_2}{\Lambda}\bar\Psi_{im}\Psi^{jm}\phi^i_{n}\phi^n_{j}+\frac{d_3}{\Lambda}\bar\Psi_{ij}\Psi^{mn}\phi^i_{m}\phi^j_{n} \, ,
    \end{split}
    \end{equation}
where $\Lambda$ is the cutoff scale, $\mgut < \Lambda \lesssim M_P$.
After setting $\phi_{24}$ to its vev,\footnote{The normalisation factor $1/\sqrt{30}$ is obtained from requiring that $\langle\phi_{24}\rangle^i_j \langle\phi_{24}\rangle^j_i = v_{\text{GUT}}^2$.}
 \begin{equation} \label{eq:phi24_vev}
        \langle\phi_{24}\rangle^i_j = \frac{1}{\sqrt{30}} \vgut \ \diag (2, 2, 2,-3,-3)\, ,
 \end{equation}
each operator gives a contribution to the dark quark masses. We obtain       
\begin{equation}
  \label{eq:mass_scalar24}
        \begin{aligned} 
            & \delta m_{\tilde D}=  \frac{1}{30}\frac{v_{\text{GUT}}^2}{\Lambda}(30c_1+4d_1)\\
            & \delta m_{\tilde L}= \frac{1}{30}\frac{v_{\text{GUT}}^2}{\Lambda}(30c_1+9d_1)\\
            & \delta m_{U}= \frac{1}{30}\frac{v_{\text{GUT}}^2}{\Lambda}(30c_2+4 d_2+4d_3)\\
            & \delta m_{E}= \frac{1}{30}\frac{v_{\text{GUT}}^2}{\Lambda}(30c_2+9d_2+9d_3)\\
            & \delta m_{Q}= \frac{1}{30}\frac{v_{\text{GUT}}^2}{\Lambda}(30c_2+\frac{13}{2}d_2-6d_3)\, .
        \end{aligned}
    \end{equation}
No operator can split naturally $Q$, $\tilde D$ from $U, E,$ and $\tilde L$.

\item $\mathbf{\phi_{75}}$:
\newline
The 75-dimensional GUT scalar field transforms as a rank-4 tensor with two anti-symmetric covariant and two anti-symmetric contravariant indices, and is traceless in pairs of upper and lower indices (see for example  Ref.~\cite{Hubsch:1984pg}).
We can thus write:
\begin{equation}
    (\phi_{75})^{ij}_{mn} =  -(\phi_{75})^{ji}_{mn} = - (\phi_{75})^{ij}_{nm}, \quad (\phi_{75})^{ij}_{in} = 0 \quad \mathrm{where} \quad i, j, ... = 1,... 5\, .
\end{equation}
The decomposition of the 75 under the $SU(3) \times SU(2) \times U(1)$ SM subgroup is:
\begin{equation}
    75 = (1,1,0) \+ (8,1 ,0) \+ (8,3,0) \+ (3,2,-5/6) \+ (\bar{6},2,-5/6) \+ (\bar{3},1,-5/3)\, .
  \end{equation}
The SM-preserving vev of $\phi_{75}$ is aligned along the sub-tensor $(1,1,0)$ and given by~\cite{Hubsch:1984pg}:~\footnote{The normalisation factor $1/\sqrt{72}$ is obtained from requiring that $\ \langle \phi_{75}\rangle^{ij}_{mn}  \langle \phi_{75}\rangle^{mn}_{ij}= s_{\text{GUT}}^2$.}
\begin{equation}\label{eq:75_vev}
        \langle
        \phi_{75}\rangle^{ij}_{mn}=\frac{s_{\text{GUT}}}{\sqrt{72}} \big[(\delta^{\alpha}_{\gamma}\delta^{\beta}_{\delta}-\delta^{\alpha}_{\delta}\delta^{\beta}_{\gamma})+3(\delta^a_c\delta^b_d-\delta^a_d\delta^b_c)-(\delta^{\alpha}_{\gamma}\delta^b_d-\delta^{\alpha}_{\delta}\delta^b_c+\delta^{\beta}_{\delta}\delta^a_c-\delta^{\beta}_{\gamma}\delta^a_d)\big],
\end{equation}
where $\sgut$ is the dimensionful parameter that characterizes the vev, and we have used the conventional index splitting given by:
\begin{equation}
  \label{eq:splitting}
\begin{aligned}
        i,j,m,n & \to (\alpha, a), (\beta, b), (\gamma, c), (\delta, d) \\[0.1cm]
        \alpha, \beta, \gamma, \delta &= 1, 2, 3 \quad \quad a, b, c, d = 4, 5\, .
\end{aligned}
\end{equation}
Here, $\alpha, \beta, \gamma, \delta$ are $SU(3)$ indices, and $a, b, c, d$ are $SU(2)$ indices. Therefore, the sub-tensor in Eq.~(\ref{eq:75_vev}) has three blocks
where $i, j, m, n$ can be respectively only $SU(3)$ indices, only $SU(2)$ indices, and mixed $SU(3)$  and $SU(2)$ indices.

There is one new Yukawa coupling and a few $D=5$ operators that can be written in terms of $\phi_{75}$ alone (Here again we use a simplified notation and denote $(\phi_{75})^{mn}_{ij} \to \phi^{mn}_{ij}$):
    \begin{equation}
    \begin{split}
        &  y'\bar\Psi_{mn}\Psi^{ij}\phi_{ij}^{mn} + \frac{f_1}{\Lambda}\bar\Psi_{i}\Psi^i\phi^{mn}_{rs}\phi_{mn}^{rs} 
        +  \frac{f_2}{\Lambda}\bar\Psi_{ij}\Psi^{ij}\phi^{mn}_{rs}\phi_{mn}^{rs} \\[0.2cm]
        + & \frac{t_1}{\Lambda}\bar\Psi_i\Psi^j\phi^{mi}_{rs}\phi_{mj}^{rs}
        +\frac{t_2}{\Lambda}\bar\Psi_{im}\Psi^{jm}\phi^{ni}_{rs}\phi_{nj}^{rs}+
        \frac{t_3}{\Lambda}\bar\Psi_{ij}\Psi^{mn}\phi^{ij}_{rs}\phi_{mn}^{rs}. 
    \end{split}
    \end{equation}
Furthermore, one can write the following mixed terms involving both $\phi_{24}$ and $\phi_{75}$:
\begin{equation}
 \frac{h_1}{\Lambda}\bar\Psi_i\Psi^j\phi^{ir}_{js}\phi_r^s+\frac{h_2}{\Lambda}\bar\Psi_{im}\Psi^{jm}\phi^{ir}_{js}\phi_r^s+\frac{h_3}{\Lambda}\bar\Psi_{ij}\Psi^{mn}\phi^{ir}_{mn}\phi^j_r\, .
\end{equation}
The corresponding corrections to the dark quark masses are:
\begin{equation}
  \label{eq:mass_scalar75}
        \begin{aligned}
            & \delta m_{\tilde{D}}=\frac{1}{72} \frac{\sgut^2}{\Lambda}(72f_1+8t_1) + \frac{\sgut \vgut}{\sqrt{30} \Lambda}10h_1\\
            & \delta m_{\T{L}}=\frac{1}{72} \frac{\sgut^2}{\Lambda}(72f_1+8t_1) -\frac{\sgut \vgut}{\sqrt{30} \Lambda}15h_1\\
            & \delta m_{U}=\frac{2y'\sgut}{\sqrt{72}} + \frac{1}{72} \frac{\sgut^2}{\Lambda}(72f_2+8t_2+4t_3) + \frac{\sgut\vgut}{\sqrt{30} \Lambda}(10h_2+4h_3)\\
            & \delta m_{E}=\frac{6y'\sgut}{\sqrt{72}} +\frac{1}{72}  \frac{\sgut^2}{\Lambda}(72f_2+8t_2+ 36t_3) + \frac{\sgut\vgut}{\sqrt{30} \Lambda}(-15h_2-18h_3)\\
            & \delta m_{Q}=\frac{-2y'\sgut}{\sqrt{72}}  + \frac{1}{72}  \frac{\sgut^2}{\Lambda}(72f_2+8t_2+ 4 t_3) + \frac{\sgut\vgut}{\sqrt{30} \Lambda}\Big(-\frac{5}{2}h_2+h_3\Big)\, .
        \end{aligned}
    \end{equation}
Again, we find that no operator can split naturally $Q$, $\tilde D$ from $U, E,$ and $\tilde L$.

\item $\mathbf{\phi_{200}}$:
  \newline
The 200-dimensional scalar field $\phi_{200}$ transforms as a rank-4 tensor with two symmetric covariant and two symmetric contravariant indices, and is traceless in pairs of upper and lower indices. Its SM-preserving vev is given by
\begin{equation}
        \langle (\phi_{200})^{ij}_{mn}\rangle= \frac{\Sgut}{\sqrt{168}}[(\delta^{\alpha}_{\gamma}\delta^{\beta}_{\delta}+\delta^{\alpha}_{\delta}\delta^{\beta}_{\gamma})+2(\delta^a_c\delta^b_d+\delta^a_d\delta^b_c)-2(\delta^{\alpha}_{\gamma}\delta^b_d+\delta^{\alpha}_{\delta}\delta^b_c+\delta^{\beta}_{\delta}\delta^a_c+\delta^{\beta}_{\gamma}\delta^a_d)]\, ,
    \end{equation} 
where we normalized the vev requiring that $\langle \phi_{200}\rangle^{ij}_{mn}  \langle \phi_{200}\rangle^{mn}_{ij}= \sigma_{\text{GUT}}^2$, and denoted with $\Sgut$ the dimensionful parameter that characterizes it. As before, we have followed the index splitting notation introduced in Eq.~(\ref{eq:splitting}).
There exist no renormalizable Yukawa interactions that can be written with $\phi_{200}$. The following $D=5$ operators involve $\phi_{200}$ alone (We use again a simplified notation and denote $(\phi_{200})^{ij}_{mn}\to \tilde \phi^{ij}_{mn}$)
 \begin{equation}
    \begin{aligned}
       & \frac{r_1}{\Lambda}\bar\Psi_i \Psi^i \tilde{\phi}^{mn}_{jk}\tilde{\phi}^{jk}_{mn}+\frac{r_2}{\Lambda}\bar\Psi_{ij}\Psi^{ij} \tilde{\phi}^{mn}_{hk}\tilde{\phi}^{hk}_{mn}+\frac{r_3}{\Lambda}\bar\Psi_i\Psi^j \tilde{\phi}^{mn}_{jk}\tilde{\phi}^{ik}_{mn}\\[0.2cm]
        &+\frac{r_4}{\Lambda}\bar\Psi_{ij}\Psi^{hj} \tilde{\phi}^{il}_{mn}\tilde{\phi}^{mn}_{hl}+\frac{r_5}{\Lambda}\bar\Psi_{ij}\Psi^{hl} \tilde{\phi}^{im}_{hn}\tilde{\phi}^{jn}_{lm}\, ,
    \end{aligned}
  \end{equation}
and other two can be written in terms of both $\phi_{24}$ and $\phi_{200}$:
\begin{equation}
        \frac{s_1}{\Lambda}\bar\Psi_i\Psi^j \phi^{h}_{k}\tilde{\phi}^{ik}_{jh}+\frac{s_2}{\Lambda}\bar\Psi_{il}\Psi^{jl} \phi^{h}_{k}\tilde{\phi}^{ik}_{jh}\, .
\end{equation}
The corresponding corrections to the dark quark masses are:
\begin{equation}
        \begin{aligned}
            &\delta m_{\tilde{D}}=\frac{1}{168} \frac{{\Sgut^2}}{\Lambda} (168r_1 + 24r_3)+ \frac{\sgut\vgut}{\sqrt{168}\sqrt{30}\Lambda}20s_1\\
            &\delta m_{\T{L}}=\frac{1}{168}  \frac{{\Sgut^2}}{\Lambda}(168r_1 + 48r_3) - \frac{\sgut\vgut}{\sqrt{168}\Lambda}\sqrt{30} s_1\\
            &\delta m_{U}=\frac{1}{168}  \frac{{\Sgut^2}}{\Lambda}(168r_2 + 24r_4 + 12r_5)+\frac{\sgut\vgut}{\sqrt{168}\sqrt{30}\Lambda}20s_2\\
            &\delta m_{E}=\frac{1}{168}  \frac{{\Sgut^2}}{\Lambda}(168r_2 + 48r_4 +24r_5) -\frac{\sgut\vgut}{\sqrt{168}\Lambda}\sqrt{30} s_2\\
            &\delta m_{Q}=\frac{1}{168} \frac{{\Sgut^2}}{\Lambda}(168r_2 + 36r_4 - 24r_5) -\frac{\sgut\vgut}{\sqrt{168}\sqrt{30}\Lambda}5s_2\, .
        \end{aligned}
    \end{equation}
In case both $\phi_{200}$ and $\phi_{75}$ are present, there exists one additional operator,
\begin{equation}
        \frac{w}{\Lambda}\bar\Psi_{ij}\Psi^{kh} \phi^{im}_{kn}\tilde{\phi}^{jn}_{hm}\, ,
\end{equation}
which gives the following contribution to the dark quark masses:
\begin{equation}
        \begin{aligned}
            &\delta m_{U}=\frac{\sgut\Sgut}{\sqrt{168}\Lambda}8w\\
            &\delta m_{E}=-\frac{\sgut\Sgut}{\sqrt{168}\Lambda} 24w\\
            &\delta m_{Q}=\frac{\sgut\Sgut}{\sqrt{168}\Lambda} 8w\, .
        \end{aligned}
    \end{equation}
Also in this case, we find that no operator can split naturally $Q$, $\tilde D$ from $U, E,$ and $\tilde L$.    
    

\end{itemize}

\bibliographystyle{JHEP}
\bibliography{bibliography}

\providecommand{\href}[2]{#2}\begingroup\raggedright\begin{thebibliography}{10}

\bibitem{Super-Kamiokande:2016exg}
{\scshape Super-Kamiokande} collaboration, K.~Abe et~al., \emph{{Search for
  proton decay via $p \to e^+\pi^0$ and $p \to \mu^+\pi^0$ in 0.31
  megaton\textperiodcentered{}years exposure of the Super-Kamiokande water
  Cherenkov detector}},
  \href{https://doi.org/10.1103/PhysRevD.95.012004}{\emph{Phys. Rev. D}
  {\bfseries 95} (2017) 012004}
  [\href{https://arxiv.org/abs/1610.03597}{{\ttfamily 1610.03597}}].

\bibitem{Giudice:2004tc}
G.~F. Giudice and A.~Romanino, \emph{{Split supersymmetry}},
  \href{https://doi.org/10.1016/j.nuclphysb.2004.08.001}{\emph{Nucl. Phys. B}
  {\bfseries 699} (2004) 65}
  [\href{https://arxiv.org/abs/hep-ph/0406088}{{\ttfamily hep-ph/0406088}}].

\bibitem{Cirelli:2005uq}
M.~Cirelli, N.~Fornengo and A.~Strumia, \emph{{Minimal dark matter}},
  \href{https://doi.org/10.1016/j.nuclphysb.2006.07.012}{\emph{Nucl. Phys. B}
  {\bfseries 753} (2006) 178}
  [\href{https://arxiv.org/abs/hep-ph/0512090}{{\ttfamily hep-ph/0512090}}].

\bibitem{Cheung:2015mea}
C.~Cheung and D.~Sanford, \emph{{Effectively Stable Dark Matter}},
  \href{https://arxiv.org/abs/1507.00828}{{\ttfamily 1507.00828}}.

\bibitem{Kilic:2009mi}
C.~Kilic, T.~Okui and R.~Sundrum, \emph{{Vectorlike Confinement at the LHC}},
  \href{https://doi.org/10.1007/JHEP02(2010)018}{\emph{JHEP} {\bfseries 02}
  (2010) 018} [\href{https://arxiv.org/abs/0906.0577}{{\ttfamily 0906.0577}}].

\bibitem{Antipin:2015xia}
O.~Antipin, M.~Redi, A.~Strumia and E.~Vigiani, \emph{{Accidental Composite
  Dark Matter}}, \href{https://doi.org/10.1007/JHEP07(2015)039}{\emph{JHEP}
  {\bfseries 07} (2015) 039}
  [\href{https://arxiv.org/abs/1503.08749}{{\ttfamily 1503.08749}}].

\bibitem{Mitridate:2017oky}
A.~Mitridate, M.~Redi, J.~Smirnov and A.~Strumia, \emph{{Dark Matter as a
  weakly coupled Dark Baryon}},
  \href{https://doi.org/10.1007/JHEP10(2017)210}{\emph{JHEP} {\bfseries 10}
  (2017) 210} [\href{https://arxiv.org/abs/1707.05380}{{\ttfamily
  1707.05380}}].

\bibitem{Contino:2018crt}
R.~Contino, A.~Mitridate, A.~Podo and M.~Redi, \emph{{Gluequark Dark Matter}},
  \href{https://doi.org/10.1007/JHEP02(2019)187}{\emph{JHEP} {\bfseries 02}
  (2019) 187} [\href{https://arxiv.org/abs/1811.06975}{{\ttfamily
  1811.06975}}].

\bibitem{Contino:2020god}
R.~Contino, A.~Podo and F.~Revello, \emph{{Composite Dark Matter from
  Strongly-Interacting Chiral Dynamics}},
  \href{https://doi.org/10.1007/JHEP02(2021)091}{\emph{JHEP} {\bfseries 02}
  (2021) 091} [\href{https://arxiv.org/abs/2008.10607}{{\ttfamily
  2008.10607}}].

\bibitem{BCV}
S.~Bottaro, R.~Contino and S.~Verma, ``{TBA}.'' Work in progress, 2024.

\bibitem{Vecchi:2021shj}
L.~Vecchi, \emph{{Axion quality straight from the GUT}},
  \href{https://doi.org/10.1140/epjc/s10052-021-09745-x}{\emph{Eur. Phys. J. C}
  {\bfseries 81} (2021) 938}
  [\href{https://arxiv.org/abs/2106.15224}{{\ttfamily 2106.15224}}].

\bibitem{Contino:2021ayn}
R.~Contino, A.~Podo and F.~Revello, \emph{{Chiral models of composite axions
  and accidental Peccei-Quinn symmetry}},
  \href{https://doi.org/10.1007/JHEP04(2022)180}{\emph{JHEP} {\bfseries 04}
  (2022) 180} [\href{https://arxiv.org/abs/2112.09635}{{\ttfamily
  2112.09635}}].

\bibitem{Ibe:2009gt}
M.~Ibe, \emph{{Small steps towards Grand Unification and the electron/positron
  excesses in cosmic-ray experiments}},
  \href{https://doi.org/10.1088/1126-6708/2009/08/086}{\emph{JHEP} {\bfseries
  08} (2009) 086} [\href{https://arxiv.org/abs/0906.4667}{{\ttfamily
  0906.4667}}].

\bibitem{Aizawa:2014iea}
T.~Aizawa, M.~Ibe and K.~Kaneta, \emph{{Coupling Unification and Dark Matter in
  a Standard Model Extension with Adjoint Majorana Fermions}},
  \href{https://doi.org/10.1103/PhysRevD.91.075012}{\emph{Phys. Rev. D}
  {\bfseries 91} (2015) 075012}
  [\href{https://arxiv.org/abs/1411.6044}{{\ttfamily 1411.6044}}].

\bibitem{Mahbubani:2005pt}
R.~Mahbubani and L.~Senatore, \emph{{The Minimal model for dark matter and
  unification}}, \href{https://doi.org/10.1103/PhysRevD.73.043510}{\emph{Phys.
  Rev. D} {\bfseries 73} (2006) 043510}
  [\href{https://arxiv.org/abs/hep-ph/0510064}{{\ttfamily hep-ph/0510064}}].

\bibitem{Harigaya:2016vda}
K.~Harigaya, T.~Lin and H.~K. Lou, \emph{{GUTzilla Dark Matter}},
  \href{https://doi.org/10.1007/JHEP09(2016)014}{\emph{JHEP} {\bfseries 09}
  (2016) 014} [\href{https://arxiv.org/abs/1606.00923}{{\ttfamily
  1606.00923}}].

\bibitem{Witten:1983tx}
E.~Witten, \emph{{Current Algebra, Baryons, and Quark Confinement}},
  \href{https://doi.org/10.1016/0550-3213(83)90064-0}{\emph{Nucl. Phys. B}
  {\bfseries 223} (1983) 433}.

\bibitem{Buttazzo:2019iwr}
D.~Buttazzo, L.~Di~Luzio, G.~Landini, A.~Strumia and D.~Teresi, \emph{{Dark
  Matter from self-dual gauge/Higgs dynamics}},
  \href{https://doi.org/10.1007/JHEP10(2019)067}{\emph{JHEP} {\bfseries 10}
  (2019) 067} [\href{https://arxiv.org/abs/1907.11228}{{\ttfamily
  1907.11228}}].

\bibitem{Palmisano:2024mxj}
S.~Palmisano, F.~Rescigno and F.~Troni, \emph{{Models of Accidental Dark Matter
  with a Fundamental Scalar}},
  \href{https://arxiv.org/abs/2403.07759}{{\ttfamily 2403.07759}}.

\bibitem{Bai:2010qg}
Y.~Bai and R.~J. Hill, \emph{{Weakly Interacting Stable Pions}},
  \href{https://doi.org/10.1103/PhysRevD.82.111701}{\emph{Phys. Rev. D}
  {\bfseries 82} (2010) 111701}
  [\href{https://arxiv.org/abs/1005.0008}{{\ttfamily 1005.0008}}].

\bibitem{Cohen:2016uyg}
T.~Cohen, K.~Murase, N.~L. Rodd, B.~R. Safdi and Y.~Soreq,
  \emph{{\ensuremath{\gamma} -ray Constraints on Decaying Dark Matter and
  Implications for IceCube}},
  \href{https://doi.org/10.1103/PhysRevLett.119.021102}{\emph{Phys. Rev. Lett.}
  {\bfseries 119} (2017) 021102}
  [\href{https://arxiv.org/abs/1612.05638}{{\ttfamily 1612.05638}}].

\bibitem{Ando:2015qda}
S.~Ando~{\,} and K.~Ishiwata, \emph{{Constraints on decaying dark matter from
  the extragalactic gamma-ray background}},
  \href{https://doi.org/10.1088/1475-7516/2015/05/024}{\emph{JCAP} {\bfseries
  05} (2015) 024} [\href{https://arxiv.org/abs/1502.02007}{{\ttfamily
  1502.02007}}].

\bibitem{Cirelli:2012ut}
M.~Cirelli, E.~Moulin, P.~Panci, P.~D. Serpico and A.~Viana, \emph{{Gamma ray
  constraints on Decaying Dark Matter}},
  \href{https://doi.org/10.1103/PhysRevD.86.083506}{\emph{Phys. Rev. D}
  {\bfseries 86} (2012) 083506}
  [\href{https://arxiv.org/abs/1205.5283}{{\ttfamily 1205.5283}}].

\bibitem{Facchinetti:2023slb}
G.~Facchinetti, L.~Lopez-Honorez, Y.~Qin and A.~Mesinger, \emph{{21cm signal
  sensitivity to dark matter decay}},
  \href{https://doi.org/10.1088/1475-7516/2024/01/005}{\emph{JCAP} {\bfseries
  01} (2024) 005} [\href{https://arxiv.org/abs/2308.16656}{{\ttfamily
  2308.16656}}].

\bibitem{Wadekar:2021qae}
D.~Wadekar and Z.~Wang, \emph{{Strong constraints on decay and annihilation of
  dark matter from heating of gas-rich dwarf galaxies}},
  \href{https://doi.org/10.1103/PhysRevD.106.075007}{\emph{Phys. Rev. D}
  {\bfseries 106} (2022) 075007}
  [\href{https://arxiv.org/abs/2111.08025}{{\ttfamily 2111.08025}}].

\bibitem{Arguelles:2019boy}
{\scshape IceCube} collaboration, C.~A. Arg\"uelles and H.~Dujmovic,
  \emph{{Searches for Connections Between Dark Matter and Neutrinos with the
  IceCube High-Energy Starting Event Sample}},
  \href{https://doi.org/10.22323/1.358.0839}{\emph{PoS} {\bfseries ICRC2019}
  (2020) 839} [\href{https://arxiv.org/abs/1907.11193}{{\ttfamily
  1907.11193}}].

\bibitem{LZ:2022lsv}
{\scshape LZ} collaboration, J.~Aalbers et~al., \emph{{First Dark Matter Search
  Results from the LUX-ZEPLIN (LZ) Experiment}},
  \href{https://doi.org/10.1103/PhysRevLett.131.041002}{\emph{Phys. Rev. Lett.}
  {\bfseries 131} (2023) 041002}
  [\href{https://arxiv.org/abs/2207.03764}{{\ttfamily 2207.03764}}].

\bibitem{DeLuca:2018mzn}
V.~De~Luca, A.~Mitridate, M.~Redi, J.~Smirnov and A.~Strumia, \emph{{Colored
  Dark Matter}}, \href{https://doi.org/10.1103/PhysRevD.97.115024}{\emph{Phys.
  Rev. D} {\bfseries 97} (2018) 115024}
  [\href{https://arxiv.org/abs/1801.01135}{{\ttfamily 1801.01135}}].

\bibitem{DeGrand:2015zxa}
T.~DeGrand, \emph{{Lattice tests of beyond Standard Model dynamics}},
  \href{https://doi.org/10.1103/RevModPhys.88.015001}{\emph{Rev. Mod. Phys.}
  {\bfseries 88} (2016) 015001}
  [\href{https://arxiv.org/abs/1510.05018}{{\ttfamily 1510.05018}}].

\bibitem{Appelquist:1996dq}
T.~Appelquist, J.~Terning and L.~C.~R. Wijewardhana, \emph{{The Zero
  temperature chiral phase transition in SU(N) gauge theories}},
  \href{https://doi.org/10.1103/PhysRevLett.77.1214}{\emph{Phys. Rev. Lett.}
  {\bfseries 77} (1996) 1214}
  [\href{https://arxiv.org/abs/hep-ph/9602385}{{\ttfamily hep-ph/9602385}}].

\bibitem{Hubsch:1984pg}
T.~Hubsch and S.~Pallua, \emph{{SYMMETRY BREAKING MECHANISM IN AN ALTERNATIVE
  SU(5) MODEL}},
  \href{https://doi.org/10.1016/0370-2693(84)91659-9}{\emph{Phys. Lett. B}
  {\bfseries 138} (1984) 279}.

\end{thebibliography}\endgroup

\end{document}